\documentclass[letterpaper,accepted=2022-01-18]{quantumarticle}
\pdfoutput=1
\usepackage{graphicx}
\usepackage{latexsym}
\usepackage{bm,array}
\usepackage{amsfonts}
\usepackage{amssymb}
\usepackage{amsmath}
\usepackage{upgreek}
\usepackage{xcolor}
\usepackage{enumerate}
\usepackage{physics}
\usepackage{array}
\usepackage{makecell}
\usepackage{colortbl}
\usepackage{hyperref}
\usepackage{underscore}
\usepackage{cite}

\newcommand{\mtx}[2]{\left(\begin{array}{#1}#2\end{array}\right)}

\newcommand{\ddd}[1]{\href{https://doi.org/#1}{doi:#1}.}

\begin{document}

%\preprint{}

\title{A quantum prediction as a collection of epistemically restricted classical predictions}


\author{William F.~Braasch Jr.}
\email[]{William.Frederick.Braasch.Jr@dartmouth.edu}
\affiliation{Department of Physics and Astronomy, Dartmouth College, Hanover, New Hampshire 03755, USA}

\author{William K.~Wootters}
\email[]{wwootter@williams.edu}
%\thanks{}
\affiliation{Department of Physics, Williams College, Williamstown, Massachusetts 01267, USA}


%\date{\today}

\begin{abstract}
Spekkens has introduced an {\em epistemically restricted
classical theory} of discrete systems,
based on discrete phase space. The 
theory manifests a number of quantum-like properties but
cannot fully imitate quantum theory because it is
noncontextual.  In this paper we show how, for a certain
class of quantum systems, the quantum description of
an experiment can be decomposed into classical 
descriptions that {are
epistemically restricted, though in a different
sense than in Spekkens' work.}
For each aspect of the
experiment---the preparation, the transformations, and
the measurement---the epistemic restriction limits
the form of the probability distribution an imagined
classical observer may use.  There are also
global constraints that the whole collection of classical
descriptions must satisfy.  Each classical description
generates its own prediction regarding the outcome
of the experiment.
%, typically weaker{---that is, less informative---}than
%the quantum prediction.  
One recovers the quantum prediction via a simple but
highly nonclassical rule: the ``nonrandom part" of the 
predicted quantum probabilities is obtained by summing the nonrandom
parts of the classically predicted probabilities. {By ``nonrandom part'' we mean
the deviation from complete randomness, that is, from
what one would expect upon measuring the fully 
mixed state.}  
\end{abstract}

% insert suggested PACS numberis s in braces on next line
%\pacs{}
%\keywords{}

\maketitle

%\section{}
%\subsection{}
%\subsubsection{}

\section{Introduction}

For any physical theory, it is generally a good thing to have 
a number of different 
formulations of the theory.  Alternative formulations
can provide novel insights and can offer practical 
problem-solving tools, and if the theory ultimately needs
to be modified, the needed modification might be more easily
identified within one formulation than within others.  

The ultimate goal motivating the present work is 
to produce an alternative
formulation of quantum theory, based on phase space rather than
Hilbert space, and, unlike the Wigner-function formulation,
based on actual probability distributions over
phase space rather than on
quasiprobability distributions (that is, distributions that
can take negative values).  
Although we do not achieve this aim in this
paper, we do succeed in re-expressing in these terms 
the
quantum theory of systems that, in the
Hilbert-space formulation, have odd-prime-power Hilbert-space
dimensions.  We discuss in Section \ref{challenges} the 
particular challenges involved in extending our work to
other dimensions.  

Part of the impetus for our work is a 2016 paper by
Spekkens based on unpublished work from 2008 \cite{Spekkens1, Spekkens2, Bartlett2012}. Spekkens presents an {\em epistemically
restricted classical theory} of discrete systems and shows that this theory
exhibits a remarkable number of quantum-like features.  By construction, however,
the theory is noncontextual and can therefore not
fully imitate quantum theory. It is also thoroughly 
discrete, with only finitely many pure states
for any given system.
{An extention of Spekkens' model designed to allow for contextuality has been introduced but is limited to the setting of a pair of toy bits \cite{Larsson2012}.}
In the present paper, we show how 
{a greater}
% at least 
part
of quantum theory can be re-expressed in terms of
classical descriptions {that are likewise epistemically restricted, though, as we explain, our restrictions are more specific than those in Spekkens' model.}
In addition to potentially
leading to a full alternative formulation of quantum
theory, we feel that this
construction provides a novel perspective on the
relation between classical and quantum physics.  

Let us use the word
``experiment'' to refer to a sequence of operations consisting
of a preparation, a transformation (or possibly a sequence
of transformations), and a measurement. 
In Spekkens' approach, an experiment performed on an 
elementary discrete system is described 
in terms of a discrete phase space that can be pictured
as a $d \times d$ array of points, where $d$ is a prime
number.  The points of this
phase space are understood to be the actual physical states of the
system---the ``ontic states.''
But an epistemic restriction
prevents an observer from knowing
the current ontic state: the most
they can know is that the system's
state lies on a particular {\em line} in the phase space.
The allowed 
transformations in Spekkens' picture are the affine-symplectic transformations, which are analogous to certain
allowed classical transformations in continuous
phase space.  Finally, the finest-grained type of measurement
allowed is a measurement that
distinguishes the lines constituting a complete set
of parallel lines, which we call a ``striation''
of the phase space.  Distinguishing one line
in such a set is analogous to
measuring, say, the position of a particle while not simultaneously
measuring its momentum or any nontrivial linear combination of
position and momentum.  (Each of these observables is 
associated with a set of parallel lines in phase space 
having a particular
slope.)

In the present paper, we similarly consider epistemically
restricted classical accounts based on a $d \times d$ phase
space with prime $d$.  (We later extend the formalism to odd
prime powers.)  Specifically, we show how to decompose
the description of a quantum experiment into a collection
of classical descriptions, each of which adheres to 
certain epistemic restrictions. As it turns out, 
these descriptions are also interdependent in that they
must satisfy a set of global constraints, which 
correspond to the constraints on the mathematical objects appearing in the standard Hilbert space formulation of quantum mechanics. 

Each of the classical descriptions of an experiment will
generate its own prediction regarding the outcome: it generates
a probability that any particular event will occur.  Much of
our focus will be on the rule by which these classical
predictions are to be combined to recover the quantum 
mechanical prediction.
Interestingly, this rule is fairly
simple, and there is a sense in which it applies uniformly to each component of our
imagined experiment, that is, to the preparation, the
transformation, and the measurement.  We will state this
rule shortly.
But first we
need to describe our discrete, epistemically restricted
classical world more precisely.

In our approach, each component of a classical experiment
is described at two
levels.  First we need to specify the ``framework''; then we
specify a probability distribution within the chosen
framework.  The framework defines the epistemic restriction that the probability distribution must adhere to. {The introduction
of the concept of a framework is perhaps the main way in which
our formalism differs from Spekkens'.}

For the preparation, the framework is a striation of the
phase space, which partitions the $d^2$ ontic
states (that is, the points of phase space) into
$d$ parallel
lines, each consisting of $d$ points. Within this framework, a classical preparation
is represented by a probability distribution over phase space that is required to be {\em uniform}
over each of the lines of the chosen striation.  Thus, as in Spekkens' model, the
most that an experimenter can know about the ontic state is that it lies on a particular line.
{(Note, though, that for Spekkens, there is an allowed epistemic state
associated with {\em each} line of phase space,
whereas for us, once a framework
has been chosen, only the lines of a specific
striation can play this role.)} We use the letter $B$ to label
a striation, because, as we explain in Section \ref{Wigner}, each striation is associated with
an orthogonal basis of the system's Hilbert space.  

In our analysis of a {\em transformation} 
(for which we will usually use the word
``channel,'' to avoid potential ambiguities), 
we begin by considering
what we call the basic ``ontic transitions.'' An ontic
transition is a transition of the system from a phase-space
point $\alpha$ to a phase-space point $\beta$. We
label this transition $\alpha \rightarrow \beta$.  
There
are $d^4$ ontic transitions.  The framework for a 
channel is labeled by a $2 \times 2$ symplectic
matrix $S$ and is a partitioning of the set of ontic
transitions into $d^2$ ``displacement classes,''
each having $d^2$ elements.  
A displacement class is the set of all ontic
transitions $\alpha \rightarrow \beta$ such that
$\beta = S\alpha + \delta$, where $\delta$ is a
phase-space vector (a displacement) that labels
the displacement class within the framework $S$.
(Here $\alpha$, $\beta$, and $\delta$ are all to be 
thought of a two-component column vectors.)
Within this framework, a classical channel
is represented by a conditional probability distribution---the probability
of $\beta$ given $\alpha$---that depends only on the displacement
class in which the transition $\alpha \rightarrow \beta$ lies.
That is, it depends only on the value of
the displacement $\delta = \beta - S\alpha$.
If we combine this conditional probability
distribution with the distribution over $\alpha$ defined 
by the preparation, it is still the case, after 
the action of the channel, that the
experimenter can know at most what line the ontic
state lies on.  
%(If an observer
%could know the value of $\beta=\alpha - S\gamma$ and
%simultaneously know the value of 
%$\beta'=\alpha - S'\gamma$ for some $S'$ not equal to $S$,
%they could figure out the ontic transition
%$\gamma \rightarrow \alpha$, and this is not
%allowed.)

Finally we consider the measurement.  For the measurement,
as for the preparation, the framework is a 
striation, which we can label $B'$.  Each outcome $E$ of the measurement
is represented by a conditional probability 
function interpreted as the probability of the outcome
$E$ when the system is at the point $\beta$.
This function
can depend only on which {\em line} of $B'$ the point $\beta$
lies on.
%The experimenter uses the classical theory to 
%compute the probability of each point $\beta$, and then uses the 
%conditional probability function to find
%the probability of the outcome $E$.  

To summarize, for an experiment consisting of a preparation,
a channel, and a measurement, a framework consists
of a striation (for the preparation), a symplectic matrix
(for the channel), and another striation (for the 
measurement).  We can label the framework by the ordered
triple $(B',S,B)$, where $B$ and $B'$ are the striations
for the preparation and measurement, respectively.  
The classical experiment itself is expressed
by a set of probability functions, each satisfying
an epistemic restriction defined by the framework. The classical theory then
yields 
the conditional probability of the particular measurement outcome for the given preparation and channel.

We should note that the ``classical theory'' consists
simply of the ordinary rules of probability.  Once we are given an initial probability
distribution over the phase space (the
preparation), and we are told with what 
probability the system at any point in phase space will move
to any other point (the channel), and with what probability
each of these points leads to the outcome $E$, it is straightforward to assign a probability to $E$. 

We now come to the question of
recovering the standard quantum prediction from
the collection of classical predictions.
It turns out that this can be done via the following 
simple rule: to recover the quantum mechanical
prediction, we combine the predictions from all
the classical experiments by
summing their ``nonrandom parts,'' that is,
their deviations from what one would have obtained
by starting with the uniform distribution over phase
space.  This sum (possibly divided by a ``redundancy
factor,'' as we will explain) is then taken to be the
nonrandom part of the quantum prediction, that is, the
deviation from what one would have obtained by starting
with the completely mixed state.
Thus it can easily happen that for each individual
classical experiment the final probability distribution
is nearly completely random, and yet the {\em quantum} prediction
is far from random.  The reader should understand, by the way,
that at this point we do not expect it
to be obvious why one should add the nonrandom
parts of the classical predictions to get
the nonrandom part of the quantum prediction.
Showing that this is indeed the case is a large part 
of what we do in this paper.  But this procedure,
though quite nonclassical,
is at least mathematically plausible: it is
related to the familiar tactic of 
%making a vector
%space out of a set of normalized objects by 
%additively shifting the
%norm to zero \cite{Scott}.
embedding the $d \times d$ density matrices in a $(d^2 - 1)$-dimensional
vector space by subtracting from each matrix the completely mixed state \cite{Scott}.

Though in this paper we begin with a standard
quantum description and decompose it into epistemically
restricted classical descriptions, we would ultimately
like to {\em begin} with classical descriptions and
show that,
under a reasonable set of requirements, they are
equivalent to a quantum description.  In a certain sense
we do this: the global constraints we identify in 
%Section \ref{lastsection} and 
Appendix D are sufficient
to guarantee that a set of classical descriptions 
will lead to a correct quantum prediction.  However, 
these constraints are more complicated than one would
like.  We hope they can ultimately be replaced by simpler rules
that emerge more naturally from our classical setting.  {In Section \ref{lastsection} we
mention one strategy by which this might be 
possible.}

One might wonder whether our work can be used
to construct an ontology of quantum theory, that is,
an answer to the question, what is really going on?
We do not see any obvious way of doing this{, though we do not rule out such an interpretation}.  
One might try to imagine each of our 
classical observers as having their own, limited
perspective on a common reality.  Then somehow
the quantum world arises from the perspectives
of all these classical observers.  
But their perspectives, 
expressed in their derived probabilities of the outcomes
of a measurement, are combined in a way that differs from
any way in which we would normally combine classical
probabilities.  It is this strange way of combining the
classical results (that is, by summing the nonrandom
parts) that makes it difficult to extract a
picture of anything that could be called ``the actual
state of affairs."

The above concepts (if not the ontology) 
will become clearer in Section \ref{mainsection}, where we show how summing the nonrandom
parts of the classical predictions leads to the
quantum prediction.
Then, in Section \ref{coherent}, we show that for most
of the classical frameworks we have considered, the nonrandom parts are in fact
{\em zero}, so that those frameworks do not contribute at all
to the quantum prediction.  
We show that the frameworks that might make a nonzero
contribution
are the ones we call ``coherent.'' This means that
the individual components of the framework---$B$,
$S$, and $B'$ for an experiment that includes
a single channel---are consistent with
each other in the sense that the symplectic 
matrix $S$ takes the original striation $B$ to
the final striation $B'$.  This consistency
also makes sense within the classical
story we are telling: our imagined classical
experimenter would presumably want to
analyze the output of
the channel in
the unique striation that actually
reveals the information
this output holds.

To prepare the ground for our work
in Sections \ref{mainsection} and \ref{coherent}, Section \ref{Wigner} reviews
three uses of {\em quasiprobabilities} in quantum
theory: (i) the discrete Wigner function,
which represents a quantum state, (ii) transition
quasiprobabilities, which represent a quantum
channel, and (iii) the quasiprobability of obtaining
a certain measurement outcome when the system is
at any given point in phase space.  Then in Section \ref{quasitoactual}, we
show how these quasiprobabilities can be used to
define the actual probability distributions that
characterize our classical experiments.  Once we
have worked out our formalism for prime dimensions and
presented an illustrative example in Section \ref{example},
we show in Section \ref{primepower} 
how to extend our
results to any system whose Hilbert-space dimension
is a power of an odd prime, such as a set of $n$
qutrits.  Powers of 2 and composite dimensions that are
not prime powers present special challenges for our
formulation; we describe these challenges in Section \ref{challenges}.
We draw conclusions in Section \ref{lastsection}.

%Given that our ultimate aim is a formulation of
%quantum theory not based on Hilbert space or
%quasiprobabilities, the reader may wonder why
%Hilbert-space concepts and quasiprobabilities
%appear so frequently in the paper.  The reason is
%simply that we wish to show how our probability
%distributions over phase-space points relate to
%familiar notions.  In our concluding Section \ref{lastsection}, we show how the theory can be expressed in a 
%self-contained way, with no reference to
%Hilbert space or quasiprobabilities,
%in the odd-prime-power case.  

As will become clear in Section \ref{quasitoactual}, our work is closely related to work
on quantum state tomography, quantum process
tomography, and quantum measurement tomography.  The probability distributions we use to represent
preparations, channels and measurement outcomes all
arise naturally in tomographic settings.  

Our work is also closely related to the
``classical'' approach to quantum theory pursued in 
Refs.~\cite{Mancini,Ibort} and other papers by the 
same authors,
in that both approaches use the tomographic representation of quantum states. 
These authors and their collaborators have built up a large body of work using this kind of representation to describe many quantum phenomena.
Although we use a similar technique for the description of states, we treat channels and measurements in a different manner that leads to an evenhanded tomographic description of all aspects of quantum operations and allows a direct comparison to classical operations. Moreover, part of our goal is to create a formulation of quantum mechanics that clearly displays the relationship between an epistemically restricted classical theory and the full quantum theory.
{The connection between different versions of Spekkens' model and the {\em stabilizer subtheory} of quantum mechanics has been thoroughly studied previously \cite{Bartlett2012, Catani2017, Catani2018}}

The goal of expressing quantum theory in terms of actual
probability distributions has motivated other authors
as well.
%Caves, 
Fuchs and Schack \cite{%Caves, 
Fuchs} have shown how to express
states, transformations, and Born's rule in terms of
genuine probability distributions, not over phase space
but over the set of outcomes of a symmetric 
informationally complete measurement.  Other researchers have
found ways of making transition probabilities 
nonnegative by adding extra structure to the discrete
phase space \cite{Cohendet1988, Cohendet1990, Hashimoto2007, Raussendorf2020}.
{
One recently studied model that does not require negativity reproduces the operational statistics of 
the $n$-qubit stabilizer formalism \cite{Lillystone2019}.
%The ontology is a combination of Pauli operators and value assignments and
%in this way, it is reminiscent of traditional proofs of contextuality.
%It bears relation to our model where the stabilizer formalism also plays a dominant role. 
%(A stabilizer state is one whose Wigner function is a uniform distribution over a line of phase space.)
That model is contextual in
the traditional sense of Kochen and Specker \cite{Kochen1967}
and Bell \cite{Bell1966} and treats the quantum
state as epistemic.
Zurel {\em et al.}~go beyond the stabilizer subtheory in introducing a nonnegative probabilistic representation that applies to a specific model of universal quantum computation
(quantum computation with magic states) \cite{Zurel2020, Zurel2021}. 
%This model, which generalizes the traditional Wigner function construction, applies only to
%and uses a generalized set of phase point operators that only 
%dimensions that are powers
%of two.
Their construction constitutes a hidden-variable model with non-unique representations of states that fits within the standard ontological models framework \cite{Harrigan2010} (it is 
measurement-noncontextual but 
preparation-contextual \cite{Spekkens2005}). 
It thus stands in contrast to our formalism in which every quantum state has a unique representation but which lies
beyond the standard ontological models framework because
of
the unusual rule for combining probabilities.  Another
difference is that we base our
formalism on a standard discrete phase space, whereas
Zurel {\em et al.}~use an expanded phase space.}
%What we believe is novel about our
%approach is the representation of a complete quantum
%experiment as a collection of classical experiments,
%along with the simple rule for obtaining the quantum 
%prediction from the collection of permitted classical %predictions.

A word about notation.  Probabilities and quasiprobabilities
play a few distinct roles in this paper.  We represent all
quasiprobabilities, including the Wigner function, by the
letter $Q$.  Probabilities involving ontic states and
ontic transitions are labeled $R$, since they are associated
with an epistemic {\em restriction}.  All probabilities
associated with genuinely observable events (as opposed 
to the unobservable ontic states and transitions) are labeled
$P$.  Finally, we use all of the expressions $Q(\cdot | \cdot)$, $R(\cdot | \cdot)$, and $P(\cdot | \cdot)$ as 
shorthand symbols for ``the probability (or quasiprobability) of [the first argument] given [the second argument],'' and not
as symbols for fixed mathematical functions.  Thus, much of
the meaning of the expression comes from the symbols inside the parentheses.

\section{Quasiprobabilities}  \label{Wigner}

Though we aim to express quantum theory in terms of 
ordinary probability distributions taking only non-negative
values, we begin with a formulation of quantum theory
based on quasiprobability distributions.  We use 
these distributions in Section \ref{quasitoactual} 
to define our non-negative probabilities.  

\subsection{States}  \label{IIA}

\begin{figure}
    \centering
    \includegraphics[width=0.45\textwidth]{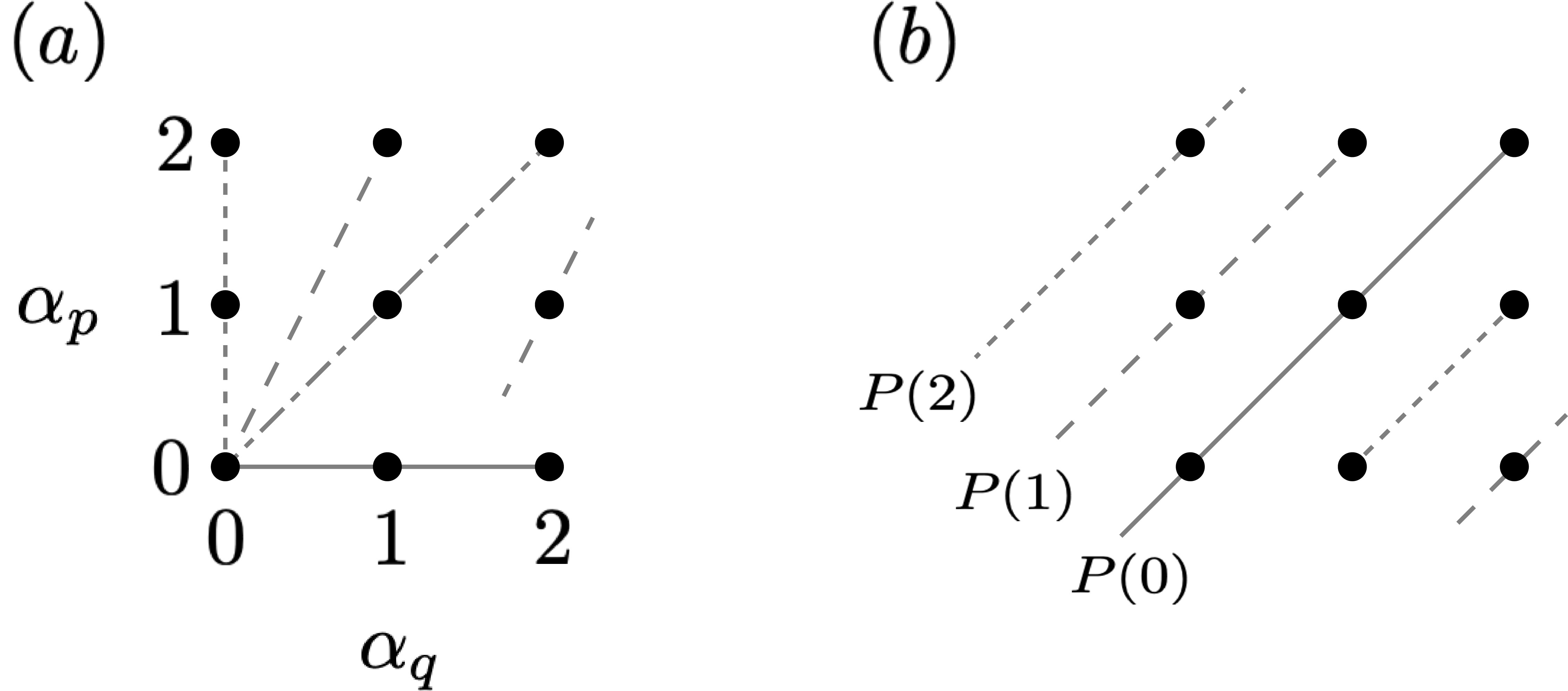}
    \caption{The discrete phase space for a qutrit is pictured as a $ 3 \times 3$ array of points. Fig.~\ref{fig:intro-phase-space}(a) shows the four 
    rays---that is, lines passing through the origin---each with different line pattern, that determine striations in the space. In Fig.~\ref{fig:intro-phase-space}(b), a choice of ray 
    with slope 1, for example, picks out a striation. Summing the Wigner function over each
    of the lines of the striation yields a marginal probability distribution. Here each different line pattern identifies a set of three points which all belong to the same phase-space line.}
    \label{fig:intro-phase-space}
\end{figure}

There are several distinct formulations of discrete phase space and discrete Wigner
functions \cite{Cohendet1988,Cohendet1990,Hashimoto2007,Buot,Hannay,Wootters,Galetti,Leonhardt1,Leonhardt2,Vourdas1,Luis,Hakioglu,Rivas,Gibbons,Vourdas2,Klimov,Vourdas4,Pittenger,Chaturvedi,Gross1,Gross2,Chaturvedi2,Vourdas3}.
For the next few sections, we use the formulation proposed
in Ref.~\cite{Wootters}, which is simplest when the
dimension $d$ of the system's Hilbert space is a prime number. In Section \ref{primepower}, where we consider prime-power
values of $d$, we use the closely related definition
proposed in Refs.~\cite{Klimov} and \cite{Vourdas4}.
For the time being, we restrict our attention to
the prime-dimensional case.

The discrete phase space for a system with (prime) dimension $d$ 
is a two-dimensional vector space over the field ${\mathbb Z}_d$, which we
picture as a $d \times d$ array of points as discussed in the
Introduction.  Phase-space points are labeled by Greek letters,
and the horizontal and vertical components of a point $\alpha$ are called $\alpha_q$ and $\alpha_p$,
respectively, in reference to the analogy with
position and momentum.  A {\em line} in this phase
space is the solution to an equation of the form
$a \alpha_q + b \alpha_p = c$, for $a,b,c \in \mathbb{Z}_d$
with $a$ and $b$ not both zero.  Two lines satisfying
linear equations that differ only in the value of $c$ are
called parallel. Fig.~\ref{fig:intro-phase-space} shows examples of lines
and parallel lines.  As we have mentioned in the Introduction,
a complete set of parallel lines is called a striation.

A general quantum state with density matrix ${w}$ is represented in this picture by its Wigner function,
defined by
\begin{equation}  \label{Wdef}
Q(\alpha|{w}) = \frac{1}{d} \hbox{tr} ({w} A_\alpha).
\end{equation}
Here the operators $A_\alpha$, called ``phase-point
operators,'' are a set of orthogonal Hermitian matrices, each with unit trace.  
For a qubit ($d = 2$), the $A$'s are defined by
\begin{equation} \label{Adefeven}
A_\alpha = \frac{1}{2} [ I + (-1)^{\alpha_p} X + (-1)^{\alpha_q + \alpha_p} Y + (-1)^{\alpha_q} Z],
\end{equation}  
where $I$ is the $2 \times 2$ identity matrix and $X$, $Y$, and $Z$ are the Pauli matrices.  
For odd $d$, the $A$'s can be defined by their matrix components:
\begin{equation}  \label{Adefodd}
(A_\alpha)_{kl} = \delta_{2\alpha_q, k+l}\, \omega^{\alpha_p (k-l)},
\end{equation}
where $\omega = e^{2 \pi i/d}$ and the matrix indices $k$ and $l$ take values
in $\mathbb{Z}_d$.   (So the addition in the subscript of the Kronecker delta is mod $d$.) 
For both even and odd $d$, the $A$'s satisfy the equation
\begin{equation} \label{Aorthogonality}
\hbox{tr}( A_\alpha A_\beta) = d \delta_{\alpha\beta},
\end{equation}
from which it follows that we can invert Eq.~(\ref{Wdef}) to express the density matrix in 
terms of the Wigner function:
\begin{equation}  \label{rhofromWigner}
{w} = \sum_\alpha Q(\alpha|{w}) A_\alpha.
\end{equation}
The Wigner function takes only real values and is normalized like a probability distribution; that is,
\begin{equation}\label{Q-rho-normalization}
\sum_\alpha Q(\alpha|{w}) = 1.
\end{equation}
But $Q(\alpha|{w})$ is not a proper probability distribution as its values can be negative.  Nevertheless, the marginal distribution
obtained by summing the Wigner function over any set of parallel lines
in phase space gives the actual probabilities
of the outcomes of an orthogonal measurement. 
(See Fig.~1(b).)
For example, the sums over the vertical lines
are the probabilities of the outcomes of a
measurement in the standard basis. This property
can be traced back to a property of the $A$
operators: for any line $\ell$,
\begin{equation}
\frac{1}{d}\sum_{\alpha \in \ell} A_\alpha = |\psi_\ell\rangle\langle \psi_\ell |,
\end{equation}
where $|\psi_\ell\rangle$ is a state vector
associated with the line $\ell$.  Because
of the orthogonality of the $A$ operators 
(Eq.~(\ref{Aorthogonality})), the state vectors associated
with two parallel lines are orthogonal.
(We will make use of this fact
in Section \ref{quasitoactual}.)

An important set of operators in the discrete
phase space picture are the 
{\em displacement} operators $D_\delta$.  They have
the property that
\begin{equation} \label{displacementproperty}
D_\delta A_\alpha D_\delta^\dag = A_{\alpha + \delta}.
\end{equation}
For $d=2$, we can write $D_\delta$ simply as
\begin{equation}
D_{\delta} = X^{\delta_q}Z^{\delta_p}.
\end{equation}
For odd $d$, we replace the Pauli matrices $X$ and $Z$ with their natural
generalizations, defined in terms of our standard orthonormal basis $\{ |m\rangle \}$ by 
\begin{equation}
X|m\rangle = |m+1\rangle \hspace{7mm} Z|m\rangle = \omega^m|m\rangle, \hspace{5mm} m \in {\mathbb Z}_d.
\end{equation}
We also adopt a convenient overall phase factor and write
\begin{equation}  \label{Dformula}
D_\delta = \omega^{\delta_q \delta_p/2} X^{\delta_q} Z^{\delta_p}.
\end{equation}
Here the division in the exponent of $\omega$
is understood to be in the field ${\mathbb Z}_d$;
for example, $\omega^{1/2} = \omega^{(d+1)/2}$.

It is often useful to write the operators $A_\alpha$ in terms of the
displacement operators.  For $d=2$, we have already done this in Eq.~(\ref{Adefeven}).
For odd $d$, the corresponding expression is
\begin{equation} \label{AfromD}
A_\alpha = \frac{1}{d} \sum_\delta \omega^{\langle \alpha, \delta \rangle} D_\delta,
\end{equation}
where $\langle \cdot, \cdot \rangle$ is the symplectic product:
\begin{equation}
\langle \alpha, \delta \rangle = \alpha_p \delta_q - \alpha_q \delta_p.
\end{equation}
Ultimately, it is the relation (\ref{AfromD}) that
gives the set of $A$'s its special structure (e.g.,
allowing the above-mentioned interpretation of
the marginals).

\subsection{Channels}  \label{IIB}

In this paper we restrict our attention to transformations of a quantum state that preserve both the dimension of the
state space and the normalization of the state.  
We refer to such transformations as
channels.
(Some authors include the dimension-preserving property
in the definition of ``channel'' as we do, while others
allow a channel to change the dimension.)  
Mathematically, a channel is represented by a completely positive, trace-preserving linear map from 
the set of operators on a Hilbert space of dimension $d$ to that same set of operators.  
Such a map can always be described by a set of $d \times d$ Kraus matrices $B_j$ satisfying 
the condition
\begin{equation}
\sum_j B_j^\dag B_j = I.
\end{equation}
In terms of these matrices, the channel can be expressed as
\begin{equation}  \label{EKrauss}
{\mathcal E}({w}) = \sum_j B_j {w} B_j^\dag .
\end{equation}

In phase space, a channel ${\mathcal E}$ is fully described by a set of transition quasiprobabilities
$Q_{\mathcal E}(\beta | \alpha)$, 
from which one can compute the new Wigner function from
the old one:
\begin{equation}  \label{newW}
Q(\beta|{\mathcal{E}({w})}) = \sum_\alpha Q_{\mathcal E}(\beta | \alpha) Q(\alpha|{w}).
\end{equation}
($Q_{\mathcal E}(\beta | \alpha)$ is an example of a Liouville
representation of a quantum channel \cite{Blum}.)
The transition quasiprobabilities representing the channel ${\mathcal E}$ are given by
\cite{Ruzzi,Zhu,Braasch}
\begin{equation}  \label{Pdeff}
Q_{\mathcal E}(\beta | \alpha) = \frac{1}{d} \hbox{tr} \left[ A_\beta {\mathcal E}(A_\alpha) \right].
\end{equation}
The normalization condition
\begin{equation}
\sum_\beta Q_{\mathcal E}(\beta | \alpha) = 1
\end{equation}
is automatically satisfied, but, like the Wigner function, the transition quasiprobabilities
can take negative values.  Thoughout this paper, we  
will assume that our channels are {\em unital}; that is,
they send the completely mixed state to itself.  
In terms of the transition quasiprobabilities, this means
that 
\begin{equation}
\sum_\alpha Q_{\mathcal E}(\beta | \alpha) = 1.
\end{equation}
So summing $Q_{\mathcal E}$ over either of its arguments
yields unity.

We now come to a fact that we rely heavily on in this paper.  There are certain linear 
transformations acting on the discrete phase space that correspond in a very simple
way to unitary transformations acting on Hilbert space \cite{Klimov, Neuhauser, Appleby, Chau}.  For such linear transformations, represented
by $2 \times 2$ matrices $S$ with entries in ${\mathbb Z}_d$, the correspondence is 
given by
\begin{equation}  \label{SUcorrespondence}
U_S A_\alpha U_S^\dag = A_{S\alpha} \;\; \hbox{for all $\alpha \in {\mathbb Z}_d^2$},
\end{equation}
where $U_S$ is a unitary transformation associated with $S$.  That is, $U_S$ bears the same relation to
the linear transformation $S$ that $D_\delta$ bears
to the displacement $\delta$. (See Eq.~(\ref{displacementproperty}).) 
If such a unitary operator exists for a given
$S$, it is unique up to an overall phase factor.
(For odd $d$, the uniqueness follows from results in Ref.~\cite{Appleby}.
For $d=2$, the proof is straightforward.) 
We will say that a transformation
$S$ for which such a unitary transformation exists is a {\em legal} linear
transformation.   
For odd $d$, the legal linear transformations are the 
symplectic transformations \cite{Appleby}, that is, the transformations
$S$ such that $S^TJS = J$, where
\begin{equation}
    J = \mtx{cc}{0 & -1 \\ 1 & 0}.
\end{equation}  
These are also the unit-determinant transformations. (Note that in a higher-dimensional phase space,
not every unit-determinant transformation is
symplectic.)  For $d=2$, the 
legal linear transformations are again symplectic, but in that case only three of the six 
symplectic matrices are legal.  The other three are equivalent to {\em antiunitary} transformations.  The {\em number}
of legal transformations, for odd $d$, is $d(d^2 - 1)$, but evidently this formula does not work 
for $d=2$.  

Even for the case of odd $d$, we will sometimes want to restrict our attention to a special set of just $d^2 - 1$ symplectic
matrices with the following property: the difference between
any two of them has nonzero determinant.  (The usefulness of this
requirement will become clear in Appendix A.)  We call such a special
set a ``minimal reconstructing set,'' because it defines the
smallest set of classical transformations that can be used
to reconstruct the quantum channel.  It is known that such
a special set exists for dimensions $2,3,5,7,$ and 11 \cite{Chau}.  To 
our knowledge, it is not yet known whether such a set exists
for any prime $d$ greater than 11.  (Chau has shown
that no set of such matrices forming a {\em group} exists for other dimensions \cite{Chau}, but we are not insisting
that the minimal reconstructing set form a group.)
Indeed, we would be happy
to use minimal reconstructing sets exclusively if it were
known that such a set exists for every prime $d$.  Though the full
set of legal matrices works well in our formalism, 
it entails a certain redundancy that we 
would avoid if the smaller sets were known to exist.
Note that for $d=2$,
the complete set of legal symplectic matrices is itself
a minimal reconstructing set.  

For convenience in the following sections, we define 
the ``redundancy factor'' ${\mathcal Z}$ to be the number
of symplectic matrices in whatever set we are using (either the
set of all legal matrices or a minimal reconstructing set) divided by $d^2 - 1$.  Thus, ${\mathcal Z}$ is given by
\begin{equation}
{\mathcal Z} = \left\{ \hspace{-2mm} \begin{array}{l}d \;\; \hbox{if $d$ is odd and we are using all legal matrices} \\1 \;\; \hbox{if $d=2$, or for any minimal reconstructing set.} \end{array} \right.
\end{equation}
This notational convention will save us from having to write separate formulas for different cases.  Specifically, we will use a factor of
$1/{\mathcal Z}$ to compensate for the redundancy 
associated with using the full set of legal matrices
when $d$ is odd.  

It is worth writing down an explicit expression for the unitary $U_S$ associated with a legal
linear transformation $S$.  For $d=2$, we label the three legal linear transformations
as ${\mathcal I}$, ${\mathcal R}$, and ${\mathcal L}$, as follows.
\begin{equation}  \label{qubitnames}
{\mathcal I} = \mtx{cc}{1 & 0 \\ 0 & 1 } \hspace{3mm} {\mathcal R}= \mtx{cc}{0 & 1 \\ 1 & 1 } \hspace{3mm}
{\mathcal L} = \mtx{cc}{1 & 1 \\ 1 & 0 }.
\end{equation}
We have chosen the last two symbols because the transformation ${\mathcal R}$ permutes the nonzero points
of phase space by rotating them to the right, that is, clockwise (for our picture of phase space), while
${\mathcal L}$ rotates them to the left.  For these matrices, the corresponding unitary transformations are
\begin{equation} 
\begin{split}
&U_{\mathcal I} = I \hspace{1.6cm}
U_{\mathcal R} = \frac{1}{\sqrt{2}}\mtx{cc}{1 & 1 \\ i & -i} \\
&U_{\mathcal L} =  \frac{1}{\sqrt{2}} \mtx{cc}{1 & -i \\ 1 & i}  .
\end{split}
\end{equation}
For odd $d$, let the linear transformation be written as
\begin{equation}
S = \mtx{cc}{v & x \\ y & z}.
\end{equation}
Then we can write the components of $U_S$ as \cite{Appleby}
\begin{equation}  \label{USformula}
(U_S)_{kl} = \left\{ \begin{array}{ll}\frac{1}{\sqrt{d}} \omega^{\frac{1}{2x}(v l^2 - 2kl + z k^2)} & \hbox{if $x \ne 0$} \\
\delta_{k, v l} \omega^{\frac{1}{2} v y l^2} & \hbox{if $x = 0$},
\end{array} \right. 
\end{equation}
where the indices $k$ and $l$ take the values 
$0, \ldots, d-1$, and again, the division in the
exponent of $\omega$ is
to be carried out in the field ${\mathbb Z}_d$.
It is possible to include additional phase factors
in the expressions for $U_S$ which guarantee that
$U_{S_1}U_{S_2} = U_{S_1 S_2}$ \cite{Neuhauser, Appleby}.  
However, in this paper we will not require this 
property and we have not included those
phase factors.

\subsection{Measurements} \label{IIC}

A general quantum measurement can be represented by 
a set of positive-semidefinite operators $E_k$ that
sum to the identity (a POVM), where $k$ labels the 
outcome.  In this paper we will typically focus on
a single outcome, which will we refer to by its
POVM operator $E$. This could be any operator satisfying
\begin{equation}
0 \le E \le I,
\end{equation}
where the latter inequality simply means that the
operator $I-E$ is positive-semidefinite.  

When considering how to represent an operator $E$ in phase space, it is useful to start with the Born rule. 
%Seeing as we are constructing representations of operators in phase space that reproduce the predictions of quantum theory, this algorithm for calculating the probability of an event $E$ given that the system is some state ${w}$ should play a central role. 
Once the representation of the state has been chosen, the Born rule picks out a representation of an event as follows. Calling upon Eq.~\eqref{rhofromWigner}, we have
\begin{equation}
\hbox{tr}(E{w}) = \sum_\alpha Q(\alpha|{w}) \hbox{tr}(E A_\alpha).
\end{equation}
It is then natural to represent $E$ by the function
\begin{equation}
Q(E|\alpha) = \hbox{tr}(E A_\alpha)
\end{equation}
so that the phase-space representation of the Born rule reads as a quasiprobabilistic law of total probability:
\begin{equation}\label{quasi-total}
P(E|{w}) = \sum_\alpha Q(E|\alpha) Q(\alpha|{w}).
\end{equation}
The function $Q(E|\alpha)$ is then interpreted as the conditional quasiprobability of observing event $E$ when
the system is at the phase-space point $\alpha$.
Although channels and measurements now both appear as conditional quasiprobability functions, their representations are rather different because of their different roles in the theory. 
The former is a map between states while the latter is inextricably enmeshed with probabilities of events.

The completely mixed state will play an important part in this work and is represented by the uniform distribution in phase space. The probability of event $E$ occurring when the system is in the completely mixed state can be calculated as 
\begin{equation}\label{prob-E-given-mixed}
P(E|I/d) = \frac{1}{d^2} \sum_\alpha Q(E|\alpha) = \frac{1}{d} \hbox{tr}(E).
\end{equation}
This is the average value of $Q(E|\alpha)$ over all of phase space.

\section{Probabilities from quasiprobabilities}  \label{quasitoactual}

In this section we show how we convert the 
quasiprobabilistic functions
of the preceding section, which can take negative
values, into non-negative functions on phase space
whose values can be interpreted as probabilities.

\subsection{States} \label{IIIA}

\subsubsection{Marginal distributions}  \label{IIIA1}

The Wigner function contains complete information about
the quantum state ${w}$.
Also containing complete information about the quantum
state is the set of {\em marginal} distributions, that is,
the sums of the Wigner function over the lines of phase
space.  

Let $B$ denote any of the $d+1$ striations of phase space.
Then the set of values
\begin{equation}  \label{PBfromQ}
P^B(\ell|{w}) = \sum_{\alpha \in \ell} Q(\alpha|{w}),
\end{equation}
for all the lines $\ell$ in the striation $B$, 
constitutes an actual, non-negative probability
distribution.  It is the distribution over the
possible outcomes of a complete orthogonal measurement
associated with $B$ when the system is in the state
${w}$.  That is,
\begin{equation}
P^B(\ell|{w}) = \langle \psi_\ell | {w} | \psi_\ell \rangle,
\end{equation}
where the states $|\psi_\ell\rangle$ with $\ell \in B$
constitute an orthogonal basis for the system's Hilbert 
space, as we discussed in Section \ref{Wigner}.  Moreover, the bases associated with distinct
striations are mutually unbiased; that is, each
vector in one basis is an equal-magnitude superposition
of the vectors in any of the other bases. (This
follows from Eq.~(\ref{Aorthogonality}) and the fact
that any two nonparallel lines intersect in exactly one
point.)

\subsubsection{Probability distributions over phase space}
\label{IIIA2}

In the description of our imagined classical
experiment, we will find it useful to have a distribution over the points of phase space, derived from the marginal distributions.
Our classical experimenter, when working in 
the framework characterized by the striation $B$, will assign
the probability $R^B(\alpha|{w})$ to the point $\alpha$, where
\begin{equation}\label{RB-eq}
R^B(\alpha|{w}) = \frac{1}{d}P^B(\ell|{w})
\end{equation}
for every point $\alpha$ lying on the line $\ell$.
That is, the probability of the line $\ell$ is
spread uniformly over that line, in keeping with the
prohibition against the classical observer having
any more detailed knowledge of the ontic state than
that it lies on a certain line.  Note that though
both $Q(\alpha|{w})$ and $R^B(\alpha|{w})$ are distributions
over phase space pertaining to the quantum state ${w}$,
they are quite distinct.  The Wigner function is 
a quasiprobability distribution containing complete
information about ${w}$, whereas $R^B(\alpha|{w})$ is a
non-negative probability distribution containing 
partial information about ${w}$.  

One might wonder why we bother at all with phase-space
{\em points}, considering that the distribution $R^B(\alpha | {w})$
contains no more information than
the distribution $P^B(\ell | {w})$.
We do this partly because the phase-space points
constitute a set of mutually exclusive possibilities
that is shared by all our imagined
classical observers.  (The $d(d+1)$ lines of phase space,
by contrast, do not constitute a set on which it
makes sense to define
a single probability distribution.)
Moreover, since the distributions $R^B$ refer to
the same sample space as the Wigner function, the conversion formula
from the $R^B$'s back to $Q$ is particularly simple, as
we now show.

\subsubsection{Reconstructing the Wigner function} \label{IIIA3}

If one has in hand all $d+1$ of the distributions
$R^B(\alpha|{w})$, or equivalently, if one has all $d+1$ of the
distributions $P^B(\ell|{w})$, one can use them to 
reconstruct the Wigner function via the following
formula:
\begin{equation}  \label{reconstructWigner}
Q(\alpha|{w}) = \frac{1}{d}\left[\sum_B P^B(\ell \ni \alpha|{w})-1\right].
\end{equation}
That is, to find the Wigner function at point $\alpha$,
we draw from each marginal distribution
the probability of the one line in that striation that contains
$\alpha$.  

Eq.~(\ref{reconstructWigner}) is a standard equation in 
quantum state tomography.  By some measures, using measurements
defined by a complete set of mutually unbiased bases is
a particularly efficient way to access the information in a quantum state \cite{WoottersFields,Petz}.  The probabilities obtained from such
measurements directly yield the Wigner function via
Eq.~(\ref{reconstructWigner}), and the density matrix
can then be obtained via Eq.~(\ref{rhofromWigner}).

We can express the content of Eq.~(\ref{reconstructWigner}) 
somewhat more simply by introducing the concept of
the ``nonrandom part'' of a probability distribution,
which we mentioned in the Introduction.  We will end up
using this concept repeatedly.  For any probability
or quasiprobability $p$,
we define $\Delta p$ to be the difference between the
value of $p$ and the value it would have had if the 
state prepared in the experiment were the completely
mixed state (represented by a uniform distribution over
phase space) and the channel were the completely
randomizing channel, that is, the channel that takes
every input state to the completely mixed state.  
Thus, for the quantities $Q(\alpha|{w})$ and $P^B(\ell|{w})$
appearing in Eq.~(\ref{reconstructWigner}), we have
\begin{equation}
\begin{split}
\Delta Q(\alpha|{w}) = Q(\alpha|{w}) - \frac{1}{d^2}, \\
\Delta P^B(\ell|{w}) = P^B(\ell|{w}) - \frac{1}{d}.
\end{split}
\end{equation}
In terms of the nonrandom parts, Eq.~(\ref{reconstructWigner})
becomes
\begin{equation}  \label{DeltaDelta}
\Delta Q(\alpha|{w}) = \frac{1}{d} \sum_B \Delta P^B(\ell \ni \alpha|{w}) = \sum_B \Delta R^B(\alpha|{w}),
\end{equation}
where $\Delta R^B(\alpha|{w}) = R^B(\alpha|{w}) - 1/d^2$,
in accordance with our general definition of the
nonrandom part.  Thus, to get the nonrandom part
of the quasiprobability distribution representing a quantum state,
we simply sum the nonrandom parts of the corresponding
classical distributions $R^B$.
We will find this form of the equation useful in 
Section \ref{mainsection}.

\subsection{Channels}  \label{IIIB}

\begin{figure*}[t]
    \centering
    \includegraphics[width=0.85\textwidth]{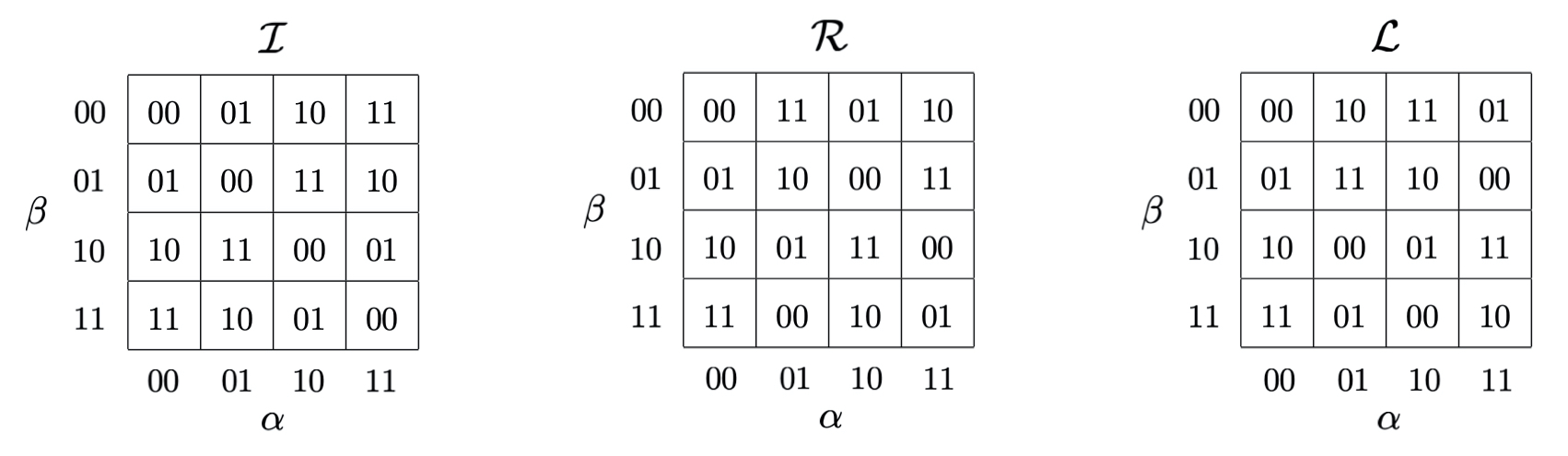}
    \caption{The three partitionings of the ontic transitions $\alpha \rightarrow \beta$ corresponding to the legal symplectic matrices for a qubit. These matrices are $\mathcal{I}$, $\mathcal{R}$, and $\mathcal{L}$ as given in Eq.~\eqref{qubitnames}. The horizontal axis, labeled $\alpha$, indicates the point of origin and the vertical axis, labeled $\beta$, indicates the point of destination. (The designation $\alpha = 01$, for example, means that
    $\alpha_q = 0$ and $\alpha_p = 1$.)  The entries 00, 01, 10, and 11 of the array indicate the displacement that is required to reach $\beta$ after the appropriate symplectic transformation is applied to $\alpha$. Note that any two displacement classes
    from different grids share exactly one ontic transition.}
    \label{fig:displacement-classes}
\end{figure*}

\subsubsection{Probabilities of displacement classes}\label{tomo}

Our treatment of quantum channels is closely analogous
to our treatment of quantum states.  A given channel
${\mathcal E}$
is completely characterized by its transition 
quasiprobabilities $Q_{\mathcal E}(\beta | \alpha)$.  Just as
a striation partitions the points of phase space
into lines, a {\em symplectic matrix}
partitions the set of ontic transitions 
$\alpha \rightarrow \beta$ into displacement classes,
as we have discussed in the Introduction. 
Fig.~2 shows the displacement classes for each of the
three legal symplectic matrices for a qubit.  

For the
symplectic matrix $S$, we define the probability
of the displacement class $\delta$ to be
\begin{equation}  \label{Thetadeff}
P^S_{\mathcal E}(\delta) = \frac{1}{d^2}\sum_{\beta,\alpha}\delta_{\beta,S\alpha + \delta}Q_{\mathcal E}(\beta | \alpha) = \frac{1}{d^2}\sum_\alpha Q_{\mathcal E}(S\alpha + \delta | \alpha).
\end{equation}
That is, we are summing over all the ontic transitions
for which the final state $\beta$ is displaced from 
$S\alpha$ by a specific vector $\delta$.  (We can think
of the factor $1/d^2$ as a uniform prior for the starting
state $\alpha$.  It is reasonable to assume this prior
when we are describing the channel on its own, 
as we are doing now.
In Subsection \ref{IIIB2} we will define a {\em conditional}
probability, which, in the context of the whole experiment,
will eventually be combined with a distribution
over $\alpha$ that need not be uniform.)
We now show that,
like the marginals discussed in Subsection \ref{IIIA1}, the
probabilities of the displacement classes are
non-negative.  We do this by showing how these probabilities
arise as the probabilities of observable events in quantum process
tomography.  

The scenario we have in mind is this.  We have an apparatus
that effects a quantum channel ${\mathcal E}$.  We will
apply the apparatus repeatedly, to identically prepared
systems, in order to fully characterize
the channel.  We use an ancilla-assisted 
method \cite{Scott,Leung1,DAriano,Dur,Leung2,Altepeter}.
That is, the system on which ${\mathcal E}$ will act in each
iteration is paired with an ancillary system
having the same Hilbert-space dimension $d$. The pair is initially prepared in the simple maximally
entangled state
\begin{equation}
|\Psi_0\rangle = \frac{1}{\sqrt{d}} \sum_k |k\rangle \otimes |k\rangle.
\end{equation}
Here we are imagining that the system on which the channel ${\mathcal E}$ will act
is the second system.  We thus apply ${\mathcal E}$ and get the state
\begin{equation}
\begin{split}
{w}_{\mathcal E} &= \sum_j (I \otimes B_j) |\Psi_0\rangle\langle \Psi_0 | (I \otimes B_j^\dag) \\
&= \frac{1}{d} \sum_{j,k,l} | k\rangle \langle l | \otimes B_j | k \rangle \langle l | B_j^\dag,
\end{split}
\end{equation}
where the $B_j$'s are a set of Kraus operators characterizing the channel ${\mathcal E}$.
Now we measure the pair in the orthonormal basis $\{ | \Phi^S_\delta \rangle \}$ defined by
\begin{equation}
| \Phi^S_\delta\rangle = \frac{1}{\sqrt{d}} \sum_m | m \rangle \otimes D_\delta U_S | m \rangle,
\end{equation}
where the index $S$ identifies the basis and the vectors in the basis
are labeled by $\delta$.  The probability of the outcome $\delta$ is
\begin{equation} \label{probsx}
\begin{split}
\langle \Phi^S_\delta | {w}_{\mathcal E} | \Phi^S_\delta \rangle
&= \frac{1}{d^2} \sum_{j,k,l} \langle k | U_S^\dag D_\delta^\dag B_j | k \rangle \langle l | B_j^\dag D_\delta
U_S | l \rangle \\
&= \frac{1}{d^2} \sum_j \left| \hbox{tr}\left( U_S^\dag D_\delta^\dag B_j \right) \right|^2.
\end{split}
\end{equation}
(That these probabilities are enough to reconstruct 
the channel $\mathcal{E}$ is shown in
Subsection \ref{IIIB3} below.)

To show that the probability in
Eq.~(\ref{probsx}) is equal to $P^S_{\mathcal E}(\delta)$, we begin by combining Eqs.~(\ref{Thetadeff}), (\ref{Pdeff}) and (\ref{EKrauss}):
\begin{equation}
\begin{split}
P^S_{\mathcal E}(\delta) &= \frac{1}{d^3} \sum_\alpha \sum_j \hbox{tr} \left( A_{S\alpha + \delta} B_j A_\alpha B_j^\dag \right)\\
&=\frac{1}{d^3} \sum_j \sum_\alpha \hbox{tr} \left( D_\delta U_S A_\alpha U_S^\dag D_\delta^\dag B_j A_\alpha B_j^\dag \right).
\end{split}
\end{equation}
Now we use the fact that for any
matrix $M$ \cite{Braasch},
\begin{equation} \label{Aequation2}
\sum_\gamma A_\gamma M A_\gamma = d(\hbox{tr} M ) I.
\end{equation}
This gives us
\begin{equation}
\begin{split}
P^S_{\mathcal E}(\delta) &= \frac{1}{d^2} \sum_j \hbox{tr}\left( U_S^\dag D_\delta^\dag B_j \right) 
\hbox{tr} \left( D_\delta U_S B_j^\dag \right) \\
&= \frac{1}{d^2} \sum_j \left| \hbox{tr} \left( U_S^\dag D_\delta^\dag B_j \right) \right|^2,
\end{split}
\end{equation}
which agrees with the set of probabilities obtained in Eq.~(\ref{probsx}).  Thus, not only is
$P_{\mathcal E}^S(\delta)$ a legitimate probability distribution; it is a distribution one can imagine
measuring directly.  

Note that in the above derivation, it is crucial that the linear transformation $S$ be
equivalent to a unitary transformation, that is, that it be ``legal.''  If we remove this 
requirement on $S$, it is not hard to find examples where $P_{\mathcal E}^S(\delta)$---still defined by
Eq.~(\ref{Thetadeff})---is negative.  

\subsubsection{Conditional distributions over phase space} \label{IIIB2}

For use later when we describe our imagined classical experiment,
we now use $P^S_{\mathcal E}(\delta)$ to define conditional probability distributions
over the points of phase space, associated with the
channel ${\mathcal E}$.  (This is analogous to our defining $R^B(\alpha|{w})$
in Subsection \ref{IIIA2}.)  We define the
probability of the system arriving at $\beta$, given that it
starts at $\alpha$, to be
\begin{equation}\label{RS-eq}
R^S_{\mathcal E}(\beta | \alpha) = P^S_{\mathcal E}(\beta - S\alpha).
\end{equation}
That is, we assume that the probability of the system
moving to point $\beta$ when it starts at point $\alpha$
is simply equal to the the probability of the
displacement class of the
transition $\alpha \rightarrow \beta$: $R^S_{\mathcal E}(\beta | \alpha)$ does not
depend in any other way on the starting point $\alpha$.
This is analogous to our assumption, in Subsection \ref{IIIA2},
that the probability $R^B(\alpha|{w})$ of a phase-space point
$\alpha$ depends only on which {\em line} $\alpha$ lies on.

{Despite this analogy, the constraint on the form of $R^S_{\mathcal E}(\beta | \alpha)$ is, in a certain respect,
less restrictive than the constraint on $R^B(\alpha | w)$.  It allows the observer within the
framework $S$ to have 
complete knowledge of certain deterministic transition rules, namely, those of the form
$\alpha \rightarrow S\alpha + \delta$ with a fixed value of $\delta$.  The set of all such transformations,
from all the frameworks $S$, is the same as the set of reversible transformations allowed in Spekkens' model \cite{Spekkens1}.}

%One might wonder whether this constraint on the form of $R^S_{\mathcal E}$ is more
%than an epistemic constraint.  There are, after all, many deterministic transitions rules
%$R^S_{\mathcal E}(\beta | \alpha) = \delta_{\beta, f(\alpha)}$ that are simply forbidden---$f(\alpha)$
%may take every phase-space point to the origin, for example.  But the constraint does not address
%the question of an underlying dynamics.  It concerns only the probabilities the observer assigns to
%events.  If all points are in fact taken to the origin, the observer cannot know this.  

%This restriction on the form of
%$R^S_{\mathcal E}$ does not place any fundamental, ontological constraint on the
%dynamics of the classical world we are imagining.  For example, the true
%transition rule for the channel $\mathcal{E}$ in this world might be the deterministic rule that takes every phase-space point to the 
%origin.  But if this is the case, the observer cannot know it.  For the observer, there must be
%other possibilities which, when figured into the observer's assignment of probabilities, give 
%$R^S_{\mathcal E}$ the form (\ref{RS-eq}).

%Knowing to which
%displacement class the ontic transition belongs exhausts the information the
%classical observer is allowed to have about the
%channel itself. 

We can understand $R^S_{\mathcal E}(\beta|\alpha)$
in another way by writing it
in terms of $Q_{\mathcal E}(\beta|\alpha)$:
\begin{equation}
R^S_{\mathcal E}(\beta|\alpha) = \frac{1}{d^2}
\sum_\mu Q_{\mathcal E}(S\mu + \delta | \mu),
\end{equation}
where $\delta = \beta - S\alpha$.  Thus, to get
the restricted probability distributions $R^S_{\mathcal E}(\beta|\alpha)$, we sum
the quasiprobability $Q_{\mathcal E}(\beta|\alpha)$ over each 
displacement class associated with $S$
and then spread the result evenly over the $d^2$
pairs $(\alpha,\beta)$ in that class,
just as, in the case of a preparation, we 
sum the Wigner function over each line in a striation and then 
spread the result evenly over that line.  

Note that 
%though both $Q_{\mathcal E}(\beta | \alpha)$ and $R^S_{\mathcal E}(\beta | \alpha)$
%are conditional phase-space distributions pertaining to the
%channel ${\mathcal E}$, they are quite distinct.  
whereas $Q_{\mathcal E}(\beta | \alpha)$ consists of quasiprobabilities and contains complete
information about ${\mathcal E}$, $R^S_{\mathcal E}(\beta | \alpha)$
is non-negative and contains only partial information
about ${\mathcal E}$.  We now show how, from a knowledge
of $R^S_{\mathcal E}(\beta | \alpha)$ for {\em all} the legal
matrices $S$, or for the matrices in a minimal reconstructing set, one can reconstruct $Q_{\mathcal E}(\beta | \alpha)$ and
thus find ${\mathcal E}$. 

\subsubsection{Reconstructing the transition quasiprobabilities} \label{IIIB3}

In this subsection we derive the following formula for
reconstructing $Q_{\mathcal E}(\beta | \alpha)$ from the set of
distributions $R^S_{\mathcal E}(\beta | \alpha)$:
\begin{equation}  \label{recoverQ}
\Delta Q_{\mathcal E}(\beta | \alpha) = \frac{1}{\mathcal Z}
\sum_S \Delta R^S_{\mathcal E}(\beta | \alpha).
\end{equation}
Recall that ${\mathcal Z}$ equals 1 for a qubit or if we
are summing over a minimal reconstructing set of symplectic matrices.
It equals $d$ if $d$ is odd and we are summing over 
all the legal symplectics.  
Here ``$\Delta$'' again indicates the nonrandom part. 
That is, 
\begin{equation}
\begin{split}
\Delta Q_{\mathcal E}(\beta | \alpha) = Q_{\mathcal E}(\beta | \alpha) - \frac{1}{d^2},\\
\Delta R^S_{\mathcal E}(\beta | \alpha) = R^S_{\mathcal E}(\beta | \alpha) - \frac{1}{d^2}.
\end{split}
\end{equation}
{So one needs to add the ``random part''
$1/d^2$ to the left-hand side of Eq.~(\ref{recoverQ})
to recover $Q$ itself.}

To derive Eq.~(\ref{recoverQ}), we begin by rewriting
Eq.~(\ref{Thetadeff}) in terms of the 
nonrandom parts:
\begin{equation}
\Delta P^S_{\mathcal E}(\delta) = \frac{1}{d^2} \sum_{\beta,\alpha}
\delta_{\beta,S\alpha+\delta} \Delta Q_{\mathcal E}(\beta | \alpha).
\end{equation}
It is helpful to insert another Kronecker delta
and a factor of $1/{\mathcal Z}$:
\begin{equation}  \label{abc}
\begin{split}
\frac{1}{\mathcal Z}\sum_{S,\delta} &\delta_{\beta, S\alpha + \delta} \Delta P^S_{\mathcal E}(\delta) \\&= \frac{1}{\mathcal{Z}d^2}
\sum_{S,\delta,\beta', \alpha'} \delta_{\beta,S\alpha+\delta}
\delta_{\beta',S\alpha' + \delta} \Delta Q_{\mathcal E}(\beta' | \alpha').
\end{split}
\end{equation}
Now we use a fact proved in Appendix A:
\begin{equation}  \label{factforappendixA}
\frac{1}{\mathcal{Z}d^2}\sum_{S,\delta}
\delta_{\beta,S\alpha+\delta}\delta_{\beta',S\alpha'+\delta}
= \delta_{\beta\beta'}\delta_{\alpha\alpha'}
+ \frac{1}{d^2}(1-\delta_{\beta\beta'} 
- \delta_{\alpha\alpha'} ).
\end{equation}
When we insert this equation into the right-hand side
of Eq.~(\ref{abc}), most of the terms yield zero
when summed over $\beta'$ and $\alpha'$.  This is because
$\sum_{\beta'}\Delta Q_{\mathcal E}(\beta'|\alpha') = 0$ (since
$Q_{\mathcal E}$ is a normalized distribution) and 
$\sum_{\alpha'}\Delta Q_{\mathcal E}(\beta'|\alpha') = 0$
(since the channel is unital).  We are left with
\begin{equation}
\frac{1}{\mathcal Z}\sum_{S,\delta} \delta_{\beta, S\alpha + \delta} \Delta P^S_{\mathcal E}(\delta)= \Delta Q_{\mathcal E}(\beta | \alpha).
\end{equation}
We can rewrite this relation as
\begin{equation}  \label{recoverQagain}
\Delta Q_{\mathcal E}(\beta | \alpha) =
\frac{1}{\mathcal Z} \sum_S \Delta P^S_{\mathcal E}(\beta - S\alpha)
=\frac{1}{\mathcal Z} \sum_S \Delta R^S_{\mathcal E}(\beta | \alpha),
\end{equation}
which is Eq.~(\ref{recoverQ}). Thus, summing the nonrandom
parts of the $R^S$'s of the restricted classical
theory---and if necessary, dividing by the redundancy
factor---yields the nonrandom part of the quantum
quasiprobability distribution $Q_{\mathcal E}$.

\subsection{Measurements}  \label{IIIC}

\subsubsection{Probability conditioned on line states} \label{IIIC1}

Again, a measurement outcome $E$ is represented in phase space
by the function 
\begin{equation}\label{meas-function}
Q(E|\alpha) = \hbox{tr}( E A_\alpha).
\end{equation}
Though this function does not have the normalization of a
quasiprobability distribution, 
it will still 
be convenient to use our ``$\Delta$'' notation
with $Q(E|\alpha)$ to
indicate that we must subtract the probability of the event $E$ given that the measured state is the maximally mixed state (cf. Eq.~\ref{prob-E-given-mixed}).
That is, we define
\begin{equation}  \label{DeltaQE}
\Delta Q(E|\alpha) = Q(E|\alpha) - \frac{1}{d}\hbox{tr} E.
\end{equation}
Or, purely in terms of phase-space concepts,
\begin{equation} 
\Delta Q(E|\alpha) = Q(E|\alpha) - \frac{1}{d^2}\sum_\mu
Q(E|\mu).
\end{equation}
Note that the sum of $\Delta Q(E|\alpha)$ over $\alpha$
is zero, just as the sum of $\Delta Q(\alpha|{w})$ is
zero for a density matrix ${w}$. 

According to Eq.~\eqref{quasi-total}, the probability $P(E|\ell)$ of getting outcome $E$ when the system is in the pure state $\ket{\psi_\ell}$ associated with the line $\ell$ is given by
\begin{equation}\label{pppp}
\begin{split}
P(E|\ell)&\equiv \langle \psi_\ell | E | \psi_\ell \rangle\\ 
&= \sum_\alpha Q(E|\alpha) Q(\alpha|\psi_\ell) \\
&= \frac{1}{d} \sum_{\alpha \in \ell} Q(E|\alpha).
\end{split}
\end{equation} 
This quantity is of course always non-negative, even though
$Q(E|\alpha)$ can be negative.

\subsubsection{Probability conditioned on a phase-space point} \label{IIIC2}

Suppose our classical experimenter is using the striation
$B$ to analyze the measurement.  For such an experimenter,
the outcome $E$ will be described by a probability
function $R^B(E | \alpha)$, interpreted as the probability
of the outcome $E$ when the system is at the phase-space
point $\alpha$. In the framework $B$, all that can matter
about the point $\alpha$ is which line it is on in the
striation $B$.  So we define
\begin{equation}\label{R-meas}
R^B(E | \alpha) = P^B(E | \ell \ni \alpha),
\end{equation}
where $P^B(E | \ell)$ is the $P(E| \ell)$ of Eq.~(\ref{pppp}) but with
the understanding that $\ell$ is chosen from the
striation $B$.  Thus $R^B(E | \alpha)$ is the average
of $Q(E|\beta)$ over all the points $\beta$ lying
on the line in $B$ that contains $\alpha$. 

\subsubsection{Recovering $Q(E|\alpha)$ from the $R^B$'s} \label{IIIC3}

We now show how to recover the function $Q(E|\alpha)$
from a knowledge of all the $R^B$'s.  This is analogous
to recovering the Wigner function from its marginals,
but the situation is somewhat different in that
$R^B(E|\alpha)$ is a probability of $E$ rather than
a probability of $\alpha$.  

To begin, for any fixed $\alpha$ let us consider the sum
\begin{equation}
\sum_B P^B(E | \ell \ni \alpha).
\end{equation}
Each term in this sum is itself an average of $d$ values
of $Q(E|\beta)$, namely, the ones with $\beta \in \ell$.
When we sum over all the lines passing through $\alpha$,
the term $Q(E|\alpha)$ appears $d+1$ times, once for each
of the $d+1$ lines passing through $\alpha$.  And any term
$Q(E|\beta)$ with $\beta \ne \alpha$ appears exactly once,
since only one line passes through both $\alpha$ and $\beta$. 
Thus we have
\begin{eqnarray}
\sum_B P^B(E | \ell \ni \alpha) &=& \frac{1}{d} \bigg(dQ(E|\alpha) + \sum_\beta 
Q(E|\beta) \bigg)\\
&=&  Q(E|\alpha) + \hbox{tr}E.
\end{eqnarray}
Solving for $Q(E|\alpha)$, we get
\begin{equation}  \label{yyyyy}
Q(E|\alpha) =  \sum_B P^B(E | \ell \ni \alpha) - \hbox{tr}E .
\end{equation}
As in Section \ref{IIIA}, the content of this equation looks 
a bit simpler if we express it in terms of the nonrandom parts.  We have defined $\Delta Q(E|\alpha)$ in
Eq.~(\ref{DeltaQE}) and according to our general rule, we have
\begin{equation}
\begin{split}
&\Delta P^B(E | \ell \ni \alpha) = 
P^B(E | \ell \ni\alpha) - \frac{1}{d}\hbox{tr}E;\\
&\Delta R^B(E | \alpha) =
R^B(E | \alpha) - \frac{1}{d}\hbox{tr}E.
\end{split}
\end{equation}
Inserting these definitions into Eq.~(\ref{yyyyy}), we get
\begin{equation} \label{Deltameas}
\Delta Q(E|\alpha) = \sum_B \Delta P^B(E|\ell \ni\alpha) = \sum_B\Delta R^B(E | \alpha).
\end{equation}
{One obtains $Q(E|\alpha)$ by adding to the left-hand
side of this equation the random part 
$(1/d)\hbox{tr}E$.}  
Note that the form of Eq.~(\ref{Deltameas}) matches that of Eq.~\eqref{DeltaDelta}. 

Eq.~(\ref{Deltameas}) provides a way of 
doing measurement tomography \cite{DAriano}, just as some of our earlier
equations are associated with state and process
tomography.  If a measuring device yields an outcome
whose mathematical description is unknown, one can
send into the device multiple copies of
systems prepared in the pure
states associated with the lines $\ell$ of phase
space.  By estimating $P(E|\ell)$, one can use
Eq.~(\ref{Deltameas}) to determine the 
function $Q(E|\alpha)$, 
which completely characterizes the measurement
outcome.

\section{The whole experiment: summing the nonrandom parts}  \label{mainsection}

We now have the necessary tools to express quantum 
probabilities for a whole experiment in terms of restricted classical
probabilities, for systems with prime-dimensional Hilbert spaces. 
The simplest quantum experiment is the preparation and subsequent measurement of a quantum state. 
For this scenario, we will explicitly show how a collection of classical experiments with our epistemic restriction can be combined to give a probabilistic prediction equivalent to the Born rule.
We will then broaden our scope to include intervening
channels.

The usual quantum mechanical description of a prepare-and-measure experiment entails the specification of a state as a density matrix ${w}$ and a measurement 
outcome as a POVM element $E$.
The Born rule can be stated using the standard operators on a Hilbert space or with quasiprobabilistic quantities (Eq.~\eqref{quasi-total}):
\begin{equation}
    P(E|{w}) \equiv \hbox{tr}[E {w}]=\sum_\alpha Q(E|\alpha) Q(\alpha|{w}).
\end{equation}
Through the relationships developed in Section \ref{quasitoactual}, the Wigner function description acts as a stepping stone towards yet another equivalent representation of the content of the Born rule. 

The first goal of the present section is to explain how the content of the Born rule can be reproduced,  {in terms of nonrandom parts}, as 
\begin{equation}\label{prep-meas-main-eq}
    \Delta P(E |{w}) = \sum_{\mathcal F}\Delta R^{\mathcal F}(E|{w}),
\end{equation}
where ${\mathcal F}$ is shorthand for the framework
$(B',B)$ of the whole experiment, and the classical
probability $R^{\mathcal F}(E | {w})$ is 
defined by
\begin{equation}  \label{RFdef}
R^{\mathcal F}(E | {w}) = \sum_\alpha R^{B'}(E|\alpha)R^B(\alpha|{w}).
\end{equation}
That is, $R^{\mathcal F}$ reflects the completely classical probabilistic reasoning that would be implemented by a classical experimenter subject to an epistemic restriction.
The claim, then, is that the nonrandom part of the quantum prediction is equal to the sum of the nonrandom parts of classical predictions
given by all the possible classical frameworks for preparations and measurements.

{Before we prove Eq.~\eqref{prep-meas-main-eq},
let us give it a more physical meaning by
interpreting $R^B(\alpha | w)$ and 
$R^{B'}(E | \alpha)$ in terms of
loss of quantum coherence.  In going from $Q(\alpha | w)$
to $R^B(\alpha | w)$, we have, in effect,
implemented a dephasing of the state $w$
in the basis $B$.  Similarly, we can get 
$R^{B'}(E | \alpha)$ from $Q(E|\alpha)$ 
by imagining a dephasing in the basis
$B'$ prior to the measurement.  If $B'$
does not match $B$, the expression 
given in
Eq.~(\ref{RFdef}) contributes nothing
to the nonrandom part in 
Eq.~\eqref{prep-meas-main-eq}.  (This claim
is justified and generalized
in Section \ref{coherent}.) So the important terms
in Eq.~\eqref{prep-meas-main-eq} are those with 
$B = B'$.  Each of these terms corresponds
to a world that is classical in the sense
that the quantum coherence in the basis $B$ has
been eliminated.  Eq.~\eqref{prep-meas-main-eq}
tells us how the predictions of the
classical experiments for all the bases
$B$ are to be combined
to produce the quantum prediction.}

We now prove Eq.~\eqref{prep-meas-main-eq}. Starting with the right-hand side of the equation, or ``$\hbox{RHS}^{E,{w}}$,'' we insert the definition
(\ref{RFdef}) and distribute the $\Delta$ into the phase-space summation as described in Appendix~\ref{distribute-Delta}:
\begin{equation}
\begin{split}
    \hbox{RHS}^{E,{w}} &= \sum_{B,B'} \Delta \left[ \sum_\alpha R^{B'}(E|\alpha)R^B(\alpha|{w}) \right] \\
    &= \sum_{B,B'}  \left[ \sum_\alpha \Delta R^{B'}(E|\alpha)\Delta R^B(\alpha|{w}) \right].
\end{split}
\end{equation}
Performing the summations over the classical frameworks used for the preparations and measurements and recalling Eqs.~\eqref{DeltaDelta}~and~\eqref{Deltameas}, we move to an expression involving only quasiprobabilities, finding
\begin{equation}
\begin{split}
    \hbox{RHS}^{E,{w}} &= \sum_\alpha \Delta Q(E|\alpha) \Delta Q(\alpha|{w})\\ 
    &= \sum_\alpha \left(Q(E|\alpha) - \frac{1}{d}\hbox{tr}(E)\right)\left(Q(\alpha|{w}) - \frac{1}{d^2}\right).
\end{split}
\end{equation}
Doing the sum and using Eqs.~\eqref{Q-rho-normalization}~and~\eqref{prob-E-given-mixed}, we have a cancellation of certain constant terms, leaving the desired result,
\begin{equation}
    \hbox{RHS}^{E,{w}} = P(E|{w}) - \frac{1}{d}\hbox{tr}(E) = \Delta P(E|{w}).
\end{equation}

It is not much more work to include a channel. In complete analogy with our result for the prepare-and-measure experiment, we will show that
\begin{equation}\label{prep-trans-meas-main-eq}
\Delta P(E | \mathcal{E},{w}) = \frac{1}{\mathcal Z}
\sum_{\mathcal F} \Delta R^{\mathcal F}(E | \mathcal{E},{w}),
\end{equation}
where ${\mathcal F}$ is now the whole framework
${\mathcal F} = (B',S,B)$ and $R^{\mathcal F}(E | \mathcal{E},{w})$ is defined to be
\begin{equation}
R^{\mathcal F}(E | \mathcal{E},{w})=
     \sum_{\beta,\alpha} R^{B'}(E|\beta)R^S_\mathcal{E}(\beta|\alpha)R^B(\alpha|{w}).
\end{equation}
Let us refer to the right-hand side of 
Eq.~(\ref{prep-trans-meas-main-eq}) as ``$\hbox{RHS}^{E,\mathcal{E},{w}}$.'' 
Again, Appendix~\ref{distribute-Delta} allows for distribution of the $\Delta$ inside the phase-space sum. The relationships in
Eqs.~\eqref{DeltaDelta},~\eqref{recoverQ},~\eqref{Deltameas} lead to
\begin{eqnarray}\label{RHS-Delta-Qs}
    \hbox{RHS}^{E,\mathcal{E},{w}} &=& \sum_{\beta,\alpha} \Delta Q(E|\beta) \Delta Q_\mathcal{E}(\beta|\alpha) \Delta Q(\alpha|{w}).
\end{eqnarray}
Unwrapping the $\Delta$ terms as a difference between the $Q$'s and the %non
random parts, we find that there are eight terms after multiplication. One of them contains three $Q$ factors. The other seven terms can be simplified using the following equations:
\begin{align}
    &\sum_\alpha Q(\alpha|{w}) = 1\\
    &\sum_\alpha Q_\mathcal{E}(\beta|\alpha) = \sum_\beta Q_\mathcal{E}(\beta|\alpha) = 1\\
    & \sum_\beta Q(E|\beta) = d\, \hbox{tr}(E).
\end{align}
The key thing to notice is that if there is an unpaired phase-space point among factors of $Q$'s, the sum over that point can be performed using one of the preceding equations.
This will leave at least one other point unpaired and in a cascading fashion, the term can be reduced to a constant of magnitude $\frac{1}{d}\hbox{tr}(E)$. The sign of the constant term is determined by how many factors of $-\frac{1}{d^2}$ or $-\frac{1}{d}\hbox{tr}(E)$ went into it. There are three positive terms and four negative terms, 
leaving us with
\begin{equation}
\begin{split}
    \hbox{RHS}^{E,\mathcal{E},{w}} &= \sum_{\beta,\alpha} Q(E|\beta) Q_{\mathcal{E}}(\beta|\alpha) Q(\alpha|{w}) - \frac{1}{d}\hbox{tr}(E)\\ 
    &= \Delta P\left(E|\mathcal{E},{w}\right),
\end{split}
\end{equation}
which is what we wished to show.

There needs to be very little tweaking of the preceding argument to generalize to any number of intervening channels between preparation and measurement. We will find that if $n$ channels are implemented,
with $\mathcal{E}_1$ acting first and $\mathcal{E}_n$ 
acting last, the
probability of the outcome $E$ is determined by
\begin{widetext}
\begin{equation} \label{prep-trans-meas-gen-eq}
\Delta P(E | \mathcal{E}_n,\ldots,\mathcal{E}_1,{w})
=\frac{1}{\mathcal{Z}^n}\sum_{\mathcal F}\Delta R^\mathcal{F}(E|
\mathcal{E}_n,\ldots,\mathcal{E}_1,{w}),
\end{equation}
where ${\mathcal F} = (B',S_n,\ldots,S_1,B)$ and
%\begin{widetext}
\begin{equation}
R^\mathcal{F}(E|
\mathcal{E}_n,\ldots,\mathcal{E}_1,{w})
= \sum_{\beta_n,\ldots,\beta_1,\alpha}R^{B'}(E|\beta_n)
R^{S_n}_{\mathcal{E}_n}(\beta_n|\beta_{n-1})
\cdots R^{S_1}_{\mathcal{E}_1}(\beta_1|\alpha)
R^B(\alpha|{w}).
\end{equation}

The only point requiring a new argument comes at the step that starts with an equation analogous to Eq.~\eqref{RHS-Delta-Qs}:
\begin{eqnarray}\label{RHS-DeltaQ-generalized}
    \hbox{RHS}^{E,\mathcal{E}_n,...,\mathcal{E}_1,{w}} = \sum_{\beta_n,...,\beta_1,\alpha} \Delta Q(E|\beta_n) \Delta Q_{\mathcal{E}_n}(\beta_n|\beta_{n-1})\cdots \Delta Q_{\mathcal{E}_1}(\beta_1|\alpha)\Delta Q(\alpha|{w}).
\end{eqnarray}
Unwrapping the $\Delta$'s and multiplying out all the factors gives $2^{n+2}$ terms. One of those is not a constant and just contains factors of $Q$'s. 
Again, because of a cascading simplification as the sum over phase-space points is performed, the other terms will all be constants of magnitude $\frac{1}{d}\hbox{tr}(E)$. The key is to count the number of terms with each sign and compare those numbers.

Let $k$ be the number of constant factors, i.e. $-\frac{1}{d^2}$ or $-\frac{1}{d}\hbox{tr}(E)$, that go into such a term. 
If $k$ is even, the term will be positive while if $k$ is odd, the term will be negative.
There will be $n+2$ choose $k$ terms having $k$ constant
factors, because we have $n+2$ factors of $\Delta Q$ in Eq.~\eqref{RHS-DeltaQ-generalized}. This means we have
%Eq.~\eqref{RHS-Delta-Qs}:
\begin{eqnarray}\label{RHS-DeltaQ-generalized2}
\hspace{-7mm}    \hbox{RHS}^{E,\mathcal{E}_n,...,\mathcal{E}_1,{w}} = \left( \sum_{\beta_n,\beta_{n-1},...,\beta_1,\alpha} Q(E|\beta_n) Q_{\mathcal{E}_n}(\beta_n|\beta_{n-1})\cdots  Q_{\mathcal{E}_1}(\beta_1|\alpha) Q(\alpha|{w})\right) + \frac{1}{d}\hbox{tr}(E)\sum_{k=1}^{n+2} (-1)^k \binom{n+2}{k}.
\end{eqnarray}
\end{widetext}
We now use an identity that follows from the binomial theorem:
\begin{equation}
 \sum_{k=1}^m (-1)^k \binom{m}{k} =  \sum_{k=0}^m (-1)^k \binom{m}{k} -1 = -1,
\end{equation}
which leads us directly to Eq~\eqref{prep-trans-meas-gen-eq}.

\section{Coherent frameworks}  \label{coherent}

For an experiment consisting of a preparation, a channel,
and a measurement, we showed in the preceding section
that
\begin{equation}  \label{bigone}
\Delta P(E|\mathcal{E},{w}) \\
= \frac{1}{\mathcal Z}\sum_{\mathcal F}\Delta R^{\mathcal F}(E|\mathcal{E},{w}),
\end{equation}
%\left[\sum_{\alpha\gamma}R^{B'}(E|\alpha)R^S_{\mathcal E}(\alpha|\gamma)R^B(\gamma|{w})\right]
where
\begin{equation}  \label{DeltaRBSB}
\Delta R^{\mathcal F}(E|\mathcal{E},{w})
= \Delta 
\Bigg[ \sum_{\beta,\alpha}R^{B'}(E|\beta)R^S_{\mathcal E}(\beta|\alpha)R^B(\alpha|{w})\Bigg].
\end{equation}
(Again, $\mathcal{F} = (B',S,B)$.)
We now show that unless $B'$ is equal to $SB$,
the expression in Eq.~(\ref{DeltaRBSB}) is zero and thus 
does not contribute to the quantum probability
$P(E|\mathcal{E},{w})$.  That is, the only frameworks
for the whole experiment that contribute to the 
probability of the outcome $E$ are the {\em coherent}
frameworks---those for which
the symplectic transformation $S$ takes the lines 
of $B$ to the lines of $B'$.

We begin by rewriting Eq.~(\ref{DeltaRBSB}) by distributing
the $\Delta$ as we have done before:
\begin{equation}
\begin{split}
\Delta R^{\mathcal{F}}(E|\mathcal{E},{w})
&=\sum_{\beta,\alpha}\Delta R^{B'}(E|\beta)
\Delta R^S_{\mathcal E}(\beta|\alpha)
\Delta R^B(\alpha | {w}).\\
&=\sum_\beta \Delta R^{B'}(E|\beta) f^{SB}(\beta),
\end{split}
\end{equation}
where
\begin{equation}
 f^{SB}(\beta)
\equiv
\sum_\alpha \Delta R^S_{\mathcal E}(\beta|\alpha)
\Delta R^B(\alpha|{w}).
\end{equation}
We now wish to show that $f^{SB}(\beta)$
is uniform over any line in the striation $SB$.  

To do this, we rewrite 
$f^{SB}$ using the definition of 
$R^S$:
\begin{equation} \label{some}
f^{SB}(\beta)
= \sum_\alpha \Delta P^S_{\mathcal E}(\beta - S\alpha)
\Delta R^B(\alpha | {w}).
\end{equation}
We want to show that the quantity in Eq.~(\ref{some})
does not change when we replace $\beta$ with a
point $\beta'$ lying on the same line as $\beta$
in the striation $SB$.  Let $\beta'$ be such a point,
and let $\delta = \beta' - \beta$.
Then we can write
\begin{equation}
f^{SB}(\beta')
= \sum_\alpha \Delta P^S_{\mathcal E}(\beta - S(\alpha-S^{-1}\delta))
\Delta R^B(\alpha | {w}).
\end{equation}
Now let $\alpha' = \alpha - S^{-1}\delta$.  Then
$\alpha$ and $\alpha'$ lie on the same line in
the striation $B$.  Therefore, since $R^B(\alpha |{w})$
is uniform over each such line, we have
\begin{equation}
\begin{split}
f^{SB}(\beta')
&= \sum_{\alpha'} \Delta P^S_{\mathcal E}(\beta - S\alpha')
\Delta R^B(\alpha' | {w})\\ &= f^{SB}(\beta).
\end{split}
\end{equation}
So $f^{SB}$ is indeed uniform over any line in 
$SB$.  

We now return to the equation
\begin{equation}  \label{PRRequation}
\Delta P(E|\mathcal{E},{w})
=\frac{1}{\mathcal Z}\sum_{B',S,B}\sum_{\beta}
\Delta R^{B'}(E|\beta)f^{SB}(\beta).
\end{equation}
Holding $B'$, $S$ and $B$ fixed for now, let us imagine
what is happening when we sum over $\beta$ if $B'$
is not the same as $SB$.  Each value of $\beta$ picks
out a unique line $\ell'$ in the striation $B'$ and 
a unique line $\ell$ in the striation $SB$.  As we
sum over $\beta$, we are in fact summing over all 
possible pairs $(\ell',\ell)$. So we can 
rewrite Eq.~(\ref{PRRequation}) as
\begin{equation}
\Delta P(E|\mathcal{E},{w})
=\frac{1}{\mathcal Z}\sum_{B',S,B}\sum_{\ell',\ell}
\Delta P^{B'}(E|\ell')g^{SB}(\ell),
\end{equation}
where $g^{SB}(\ell)=f^{SB}(\alpha \in \ell)$.
The sums over $\ell'$ and $\ell$ are now
independent of each other, and the first of these sums
has the value zero because of the way $\Delta$ is
defined.  So any triple $(B',S,B)$ for which 
$B'$ is different from $SB$ does not contribute
to $P(E|\mathcal{E},{w})$.  

An iterated version of this argument 
works when there is a sequence of channels
$\mathcal{E}_n,\ldots,\mathcal{E}_1$: for a 
framework $(B',S_n, \ldots, S_1, B)$ to contribute, we must have
$B' = S_n \cdots S_1B$.

As we noted in the Introduction, it makes sense that
our imagined classical experimenter, in choosing a
framework, would want to choose a framework for the
measurement that ``matched'' the framework for the rest
of the experiment.  Any striation other than
$B' = S_n\ldots S_1 B$ would make for a measurement
incapable of revealing any information about what
had gone before.  Indeed, we could have insisted from
the beginning that what we {\em mean} by a framework for
the experiment is a coherent framework.  We have shown
in this section that this approach yields exactly the
same result as one obtains with the much larger sum,
over all values of $B',S_n,\ldots,S_1,B$ and with no 
constraint on the relation among these elements.
The larger sum simply includes many terms that are
equal to zero.

\section{An example}  \label{example}

\begin{figure*}[t]
    \centering
    \includegraphics[width=1.0\textwidth]{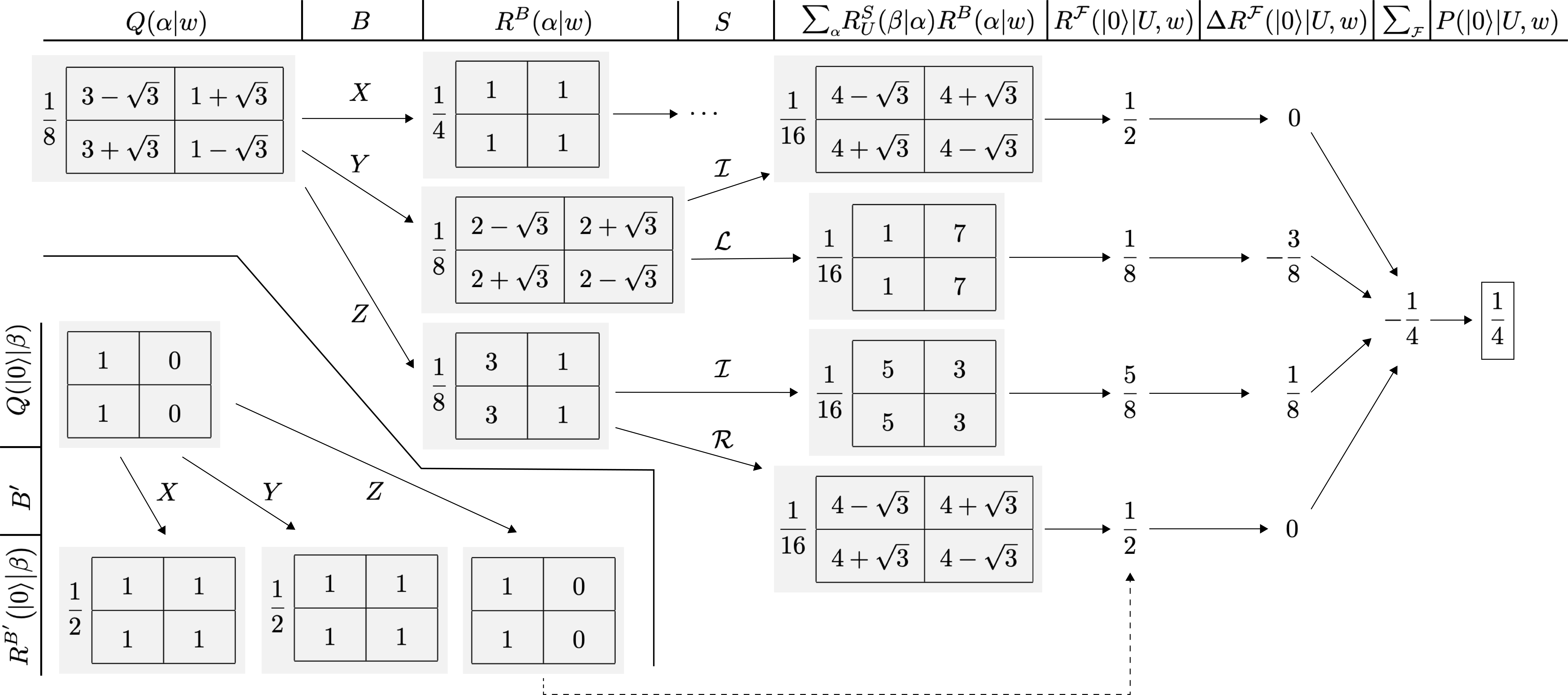}
    \caption{The steps leading from quantum distributions ($Q$'s) to epistemically restricted classical probability distributions ($R$'s), and then to a final probabilistic prediction (P). The top line contains labels for the objects contained in imagined columns below. Under these labels, and starting in the top left with the Wigner function for a specific state, we follow the flow of arrows corresponding to specific choices of classical frameworks until we reach the only nonuniform premeasurement classical states.
    These lie in the column labeled $\sum_\alpha R_U^S(\beta|\alpha) R^B(\alpha|{w})$.
    Separately, the bottom left section shows how quantum measurement distributions are related to epistemically resricted measurement distributions by a choice of measurement framework. %($X$, $Y$, or $Z$). 
    The measurement distribution consistent with a choice of classical measurement framework $Z$ leads to a sum over the left column of the premeasurement state. This choice is indicated with the dashed arrow at the bottom. 
    A sum of the nonrandom parts of the epistemically restricted classical probabilities ($\Delta R^\mathcal{F}(\ket{0}|U,{w})$'s) added to the ``random part'' ($1/2$) gives a probabilistic prediction in agreement with the Born rule. Note that there are many frameworks leading to the uniform distribution, as indicated by the ``$\cdots$'' in the fourth column, that have no effect on the final prediction.}
    \label{fig:M-Z-example}
\end{figure*}

As a simple example, consider a photon that has passed through
the first beamsplitter of a Mach-Zehnder interferometer and
is about to pass through the second beamsplitter before
being detected. States of light in the two possible paths
are labeled $\ket{0}$ and $\ket{1}$. 
The beamsplitters are unbalanced such that each has a probability $3/4$ of direct transmission (which preserves
the value of $j$ in $|j\rangle$) and $1/4$ of reflection (which
flips $j$). Specifically, let us take the unitary transformation representing the beamsplitter to be 
\begin{equation}
U = \frac{\sqrt{3}}{2}I + \frac{i}{2}X.
\end{equation}
The state prepared by the first beamsplitter is ${w} = \frac{\sqrt{3}}{2} \ket{0} + \frac{i}{2} \ket{1}$.
The Born rule gives a probability of $1/4$ that the photon will be found in the path $\ket{0}$ after the second beamsplitter. 

This section contains a careful walk-through of how the collection of classical experiments we have described above will reproduce this prediction.
Fig.~\ref{fig:M-Z-example} shows how one might visualize the steps leading from quasiprobabilities to classical probabilities and then finally to a prediction in agreement with the Born rule. All these steps in the figure are described in what follows.

This example allows us to display the mechanics of constructing the classical entities from the quantum ones. 
One can see how quasiprobabilities with distinctly quantum negative values are rendered classical as positive probability distributions at each step of the process. Negativity then returns as we sum the nonrandom parts but this is always made positive by adding in the uniform distribution at the end.

The initial superposition state has a discrete Wigner function
\begin{equation}  \label{Q-prepared}
Q(\alpha|{w}) = \frac{1}{8} \;
\bgroup 
\def\arraystretch{1.7} 
\begin{tabular}{|c|c|}
\hline
$\;\; 3-\sqrt{3} \;\;$ & $\;\; 1+\sqrt{3} \;\;$ \\
\hline
$3+\sqrt{3}$ & $\;\; 1-\sqrt{3} \;\;$   \\
\hline
\end{tabular}
\egroup \; .
\end{equation}
Note the sign of the value at the phase space point $10$ (read as ``$(1,0)$''---bottom right). 
We proceed towards the classical representation by calculating the marginal probabilities over the horizontal, diagonal, and vertical striations that we label with $X$, $Y$, and $Z$, respectively. 
The two lines in each striation will be labeled with \hbox{$x=0$} and $x=1$, where all lines that intersect the origin (bottom left) are labeled 0. 
The marginal probability distributions for the different striations are
\begin{equation}
    \begin{split}
& P^X(x) = \begin{cases}
\frac{1}{2} &\hbox{if } x = 0\\ \frac{1}{2} &\hbox{if } x = 1,
\end{cases}\\
& P^Y(x) = \begin{cases}
\frac{1}{4}(2+\sqrt{3}) &\hbox{if } x = 0\\ \frac{1}{4}(2-\sqrt{3}) &\hbox{if } x = 1,
\end{cases}\\
& P^Z(x) = \begin{cases}
\frac{3}{4} &\hbox{if } x = 0\\ \frac{1}{4} &\hbox{if } x = 1,
\end{cases}
    \end{split}
\end{equation}
Any single classical experimenter may have knowledge of at most one of these distributions. With each $P$ value spread
uniformly over its line, the epistemic state of knowledge given in Eq.~\eqref{RB-eq} will be one of the following:
\begin{equation}
\begin{split}
R^X(\alpha|{w}) &= \frac{1}{4} \;
\bgroup 
\def\arraystretch{1.7} 
\begin{tabular}{|c|c|}
\hline
$\;\; 1 \;\;$ & $\;\; 1 \;\;$ \\
\hline
$1$ & $\;\; 1 \;\;$   \\
\hline
\end{tabular}
\egroup \; ,\\
R^Y(\alpha|{w}) &= \frac{1}{8} \;
\bgroup 
\def\arraystretch{1.7} 
\begin{tabular}{|c|c|}
\hline
$\;\; 2-\sqrt{3} \;\;$ & $\;\; 2+\sqrt{3} \;\;$ \\
\hline
$2+\sqrt{3}$ & $\;\; 2-\sqrt{3} \;\;$   \\
\hline
\end{tabular}
\egroup \; ,\\
R^Z(\alpha|{w}) &= \frac{1}{8} \;
\bgroup 
\def\arraystretch{1.7} 
\begin{tabular}{|c|c|}
\hline
$\;\; 3 \;\;$ & $\;\; 1 \;\;$ \\
\hline
$3$ & $\;\; 1 \;\;$   \\
\hline
\end{tabular}
\egroup \; .
\end{split}
\end{equation}
The classical transformations allowed in the experiments are all stochastic combinations of permutations (the affine symplectic transformations) so the uniform distribution will always be mapped to itself. 
We are only interested in the nonrandom parts of distributions so $R^X$ will have no effect on the calculation going forward and can be ignored.

Moving on to the classical transformations associated with the unitary operation specified above, we start with the transition quasiprobabilities that contain full information about the quantum channel. Employing Eq.~\eqref{Pdeff}, and treating $Q_U(\beta|\alpha)$ as a matrix with $\beta$ labeling the rows and $\alpha$ labeling the columns (both ordered 00, 01, 10, 11), we have
\begin{equation}
    Q_U = \frac{1}{4}
    \begin{pmatrix}
    1 & \sqrt{3} & 3 & -\sqrt{3} \\
    -\sqrt{3} & 1 & \sqrt{3} & 3 \\
    3 & -\sqrt{3} & 1 & \sqrt{3} \\
    \sqrt{3} & 3 & -\sqrt{3} & 1
    \end{pmatrix} \; .
\end{equation}
From here, the allowed ontic transitions in a specific classical framework can be calculated using Eq.~\eqref{RS-eq} where we treat $R_\mathcal{E}^S(\beta|\alpha)$ as a matrix with rows labelled with $\beta$ and columns with $\alpha$.
For the three legal symplectic transformations $\mathcal{I}$, $\mathcal{R}$, and $\mathcal{L}$, we find
\begin{equation}
\begin{split}
R^\mathcal{I}_U &= \frac{1}{4}
\begin{pmatrix}
1 & 0 & 3 & 0 \\
0 & 1 & 0 & 3 \\
3 & 0 & 1 & 0 \\
0 & 3 & 0 & 1
\end{pmatrix},\\
R^\mathcal{R}_U &= \frac{1}{8}
\begin{pmatrix}
2 + \sqrt{3} & 2 + \sqrt{3} & 2 - \sqrt{3} & 2 - \sqrt{3} \\
2 - \sqrt{3} & 2 - \sqrt{3} & 2 + \sqrt{3} & 2 + \sqrt{3} \\
2 - \sqrt{3} & 2 - \sqrt{3} & 2 + \sqrt{3} & 2 + \sqrt{3} \\
2 + \sqrt{3} & 2 + \sqrt{3} & 2 - \sqrt{3} & 2 - \sqrt{3}
\end{pmatrix},\\
R^\mathcal{L}_U &= \frac{1}{8}
\begin{pmatrix}
2 - \sqrt{3} & 2 + \sqrt{3} & 2 + \sqrt{3} & 2 - \sqrt{3} \\
2 - \sqrt{3} & 2 + \sqrt{3} & 2 + \sqrt{3} & 2 - \sqrt{3} \\
2 + \sqrt{3} & 2 - \sqrt{3} & 2 - \sqrt{3} & 2 + \sqrt{3} \\
2 + \sqrt{3} & 2 - \sqrt{3} & 2 - \sqrt{3} & 2 + \sqrt{3}
\end{pmatrix}.
\end{split}
\end{equation}
In a single classical experiment, a transformed epistemic state is given by 
\begin{equation}
\sum_\alpha R^S_U(\beta|\alpha) R^B(\alpha|{w}).
\end{equation}
This calculation is conveniently accomplished by using the matrix form of $R^S_U(\beta|\alpha)$ and a vectorized form of $R^B(\alpha|{w})$. 
Here are the only nonuniform distributions for transformed states a classical experimenter may find depending on a choice of framework:
\begin{equation}\label{premeas-epistemic}
\begin{split}
R^\mathcal{I} R^Y &=\frac{1}{16} \;
\bgroup 
\def\arraystretch{1.7} 
\begin{tabular}{|c|c|}
\hline
$\;\; 4-\sqrt{3} \;\;$ & $\;\; 4+\sqrt{3} \;\;$ \\
\hline
$4+\sqrt{3}$ & $\;\; 4-\sqrt{3} \;\;$   \\
\hline
\end{tabular}
\egroup \; ,
\\
R^\mathcal{L} R^Y &=\frac{1}{16} \;
\bgroup 
\def\arraystretch{1.7} 
\begin{tabular}{|c|c|}
\hline
$\;\; 1 \;\;$ & $\;\; 7 \;\;$ \\
\hline
$1$ & $\;\; 7\;\;$   \\
\hline
\end{tabular}
\egroup \; ,
\\
R^\mathcal{I} R^Z &=\frac{1}{16} \;
\bgroup 
\def\arraystretch{1.7} 
\begin{tabular}{|c|c|}
\hline
$\;\; 5 \;\;$ & $\;\; 3 \;\;$ \\
\hline
$5$ & $\;\; 3\;\;$   \\
\hline
\end{tabular}
\egroup \; ,
\\
R^\mathcal{R} R^Z &=\frac{1}{16} \;
\bgroup 
\def\arraystretch{1.7} 
\begin{tabular}{|c|c|}
\hline
$\;\; 4-\sqrt{3} \;\;$ & $\;\; 4+\sqrt{3} \;\;$ \\
\hline
$4+\sqrt{3}$ & $\;\; 4-\sqrt{3} \;\;$   \\
\hline
\end{tabular}
\egroup \; .
\end{split}
\end{equation}
We now have the full collection of final states an experimenter may find. Remember that there are many unlisted ones that result in the uniform distribution. 

We had set out to calculate the probability of a photon being found in path $|0\rangle$ of the interferometer after the beamsplitter. 
In our formalism, this outcome is equivalent to the experimenter choosing the measurement framework $B' = Z$ corresponding to the vertical striation and finding the system in the left line of that striation. 
This is because the effect $E = \ket{0}\bra{0}$ is represented by the measurement function given by Eq.~\eqref{meas-function}:
\begin{equation}  \label{Q-meas}
Q({E}|\beta) =
\bgroup 
\def\arraystretch{1.7} 
\begin{tabular}{|c|c|}
\hline
$\;\; 1 \;\;$ & $\;\; 0 \;\;$ \\
\hline
$1$ & $\;\; 0 \;\;$   \\
\hline
\end{tabular}
\egroup \; .
\end{equation}
We calculate $R^B(E|\beta)$ for the different frameworks using Eq.~\eqref{R-meas} leading to
\begin{equation}
    \begin{split}
R^X({E}|\beta) &= \frac{1}{2} \;
\bgroup 
\def\arraystretch{1.7} 
\begin{tabular}{|c|c|}
\hline
$\;\; 1 \;\;$ & $\;\; 1 \;\;$ \\
\hline
$1$ & $\;\; 1 \;\;$   \\
\hline
\end{tabular}
\egroup \; ,\\
R^Y({E}|\beta) &= \frac{1}{2} \;
\bgroup 
\def\arraystretch{1.7} 
\begin{tabular}{|c|c|}
\hline
$\;\; 1 \;\;$ & $\;\; 1 \;\;$ \\
\hline
$1$ & $\;\; 1 \;\;$   \\
\hline
\end{tabular}
\egroup \; ,\\
R^Z({E}|\beta) &= 
\quad %\; \; \;
\bgroup 
\def\arraystretch{1.7} 
\begin{tabular}{|c|c|}
\hline
$\;\; 1 \;\;$ & $\;\; 0 \;\;$ \\
\hline
$1$ & $\;\; 0 \;\;$   \\
\hline
\end{tabular}
\egroup \; .
\end{split}
\end{equation}
When one calculates the classical predictions given by an equation such as
\begin{equation}
    R^{B',S,B}(E|U,{w}) = \sum_{\beta,\alpha} R^{B'}(E|\beta)R^S_\mathcal{E}(\beta|\alpha)R^B(\alpha|{w}),
\end{equation}
the choice of $B' = Z$ clearly results in the sum over the points in left vertical line of the premeasurement epistemic state (such as those given in Eq.~\eqref{premeas-epistemic}). 
Meanwhile, a choice of $B' = X \hbox{ or } Y$ will always result in a probability of $1/2$ for $E = \ket{0}\bra{0}$.

The marginal probabilities for the left vertical line of each phase space distribution in Eqs.~\eqref{premeas-epistemic} are 1/2, 1/8, 5/8,  and 1/2, respectively. We sum the nonrandom parts of these to get $0 - 3/8 + 1/8 +0 = -1/4$. Adding the 
%uniform --- I changed this wod
{random} part back in, we have $-1/4 + 1/2 = 1/4$; this is the probability of the photon being measured in path $|0\rangle$ as predicted by the collection of classical experiments. By construction, it is consistent with the Born rule.

\section{Prime-power dimensions}  \label{primepower}

Consider a set of $n$ basic systems---we will call them
particles---each having odd prime Hilbert-space
dimension $r$.  So $d=r^n$.  For the phase space
of this system, we could consider using the
$2n$-dimensional vector space over ${\mathbb Z}_r$.
However, we find it more convenient to use the
two-dimensional vector space over the finite field with
$d$ elements, 
${\mathbb F}_d$.  (Note that for $n>1$, the elements
of this field cannot be identified with ordinary 
integers.  They are more abstract objects, with
addition and multiplication rules satisfying the
requirements of a field.)  A point $\alpha$ in this phase
space still has two components $\alpha_q$ and $\alpha_p$,
but each of these components is now an element of
${\mathbb F}_d$.  The concepts of ``line'' and 
``parallel lines'' are defined exactly as in
Section \ref{Wigner}, and because of the field
properties, it is still the case that any two
nonparallel lines intersect in exactly one point.
As before, there are $d+1$ striations, each consisting of
$d$ parallel lines.  

It is worth reviewing a few features of finite
fields \cite{Lidl}.  First, a {\em basis} for the field
${\mathbb F}_{r^n}$ is an ordered set of $n$ field
elements $\{e_j\}$ such that any field element $x$ can
be written as a linear combination of the $e_j$'s
with coefficients in ${\mathbb F}_r$; that is, 
$x = \sum_j x_j e_j$ with $x_j \in {\mathbb F}_r$.
(When $r$ is prime, the field ${\mathbb F}_r$ is the
same as ${\mathbb Z}_r$.)
If we represent a field element $x$ as a column vector
with entries $x_j$, then multiplication by any field
element can be represented as ordinary matrix multiplication.
In this way, we can also represent any field element
$x$ as a matrix $M_x$ with entries in ${\mathbb F}_r$.
The {\em trace} of this matrix is independent of the 
chosen basis and is called the trace of $x$:
\begin{equation}  \label{trace2}
\hbox{Tr}_{\mathbb{F}_d/\mathbb{F}_r}(x) = \hbox{tr}\,M_x,
\end{equation}
where we use the subscript ${\mathbb{F}_d/\mathbb{F}_r}$ to
indicate that the field trace is a function from 
$\mathbb{F}_d$ to $\mathbb{F}_r$.  In what follows
we will use ``Tr'' in place of 
``$\hbox{Tr}_{\mathbb{F}_d/\mathbb{F}_r}$''
until the final paragraph of this section, where
we need to distinguish two field traces. The trace
is a linear function: (i)
For any $x \in {\mathbb F}_d$ and $a \in {\mathbb F}_r$,
$\hbox{Tr}(ax) = a\,  \hbox{Tr}(x)$. (ii) For any
$x,y \in {\mathbb F}_d$, $\hbox{Tr}(x+y) = \hbox{Tr}(x)
+ \hbox{Tr}(y)$.  The trace can also be expressed as
\begin{equation}  \label{trace1}
\hbox{Tr}(x) = x + x^r + x^{r^2} + \cdots + x^{r^{n-1}},
\end{equation}
which avoids any mention of a basis.
Finally, for every basis $\{e_j\}$,
there is a unique {\em dual basis} $\{\tilde{e}_j\}$, defined by
the property that
\begin{equation}
\hbox{Tr}(\tilde{e}_i e_j) = \delta_{ij}.
\end{equation}

We use the expansion in a basis to write a phase-space
coordinate such as $\alpha_q$ in a way that gives each
of the $n$ particles its own phase-space variable.  
Let us choose a basis $\{e_j\}$ to associate
with the horizontal component $\alpha_q$.  In the
expansion
\begin{equation}
\alpha_q = \alpha_{q1}e_1 + \cdots + \alpha_{qn}e_n,
\end{equation}
we think of $\alpha_{qj}$ (which is in ${\mathbb F}_r$)
as the horizontal phase-space coordinate of the 
$j$th particle.  For the vertical component $\alpha_p$,
we use the dual basis:
\begin{equation}
\alpha_p = \alpha_{p1}\tilde{e}_1 + \cdots + \alpha_{pn}\tilde{e}_n.
\end{equation}
The reason for this choice will become evident
shortly. 

As the Wigner function for this system, we use the
definition proposed by Vourdas \cite{Vourdas4} %Vourdas5}
and by Klimov and Muñoz \cite{Klimov}. The phase-point operators look very much
like the ones defined in Eq.~(\ref{Adefodd}), but
with a field trace in the exponent of $\omega$, where $\omega$ is now equal to
$e^{2 \pi i/r}$:
\begin{equation} \label{Adeffield}
(A_\alpha)_{kl} = \delta_{2\alpha_q, k+l}\, \omega^{\hbox{\scriptsize Tr}[\alpha_p (k-l)]}.
\end{equation}
Here $k$ and $l$, like $\alpha_q$ and $\alpha_p$,
are elements of ${\mathbb F}_d$.  The Wigner
function is defined in the same way as before
in terms of the phase-point operators:
\begin{equation}
Q(\alpha|{w}) = \frac{1}{d}\hbox{tr}({w} A_\alpha).
\end{equation}

We can understand the phase-point operators better
by expanding the phase-space coordinates in their
respective bases.  Since $k$ and $l$ label the vectors
of the
standard basis, which is associated with
the horizontal coordinate, we expand $k$ and $l$
in the basis $\{e_j\}$ associated with that 
coordinate.  In Eq.~(\ref{Adeffield}), because $(k-l)$ is expanded in the
basis $\{e_j\}$ while $\alpha_p$ is expanded in 
the dual basis $\{\tilde{e}_j\}$, the exponent
of $\omega$ becomes simply
\begin{equation}
\alpha_{p1}(k_1-l_1) + \cdots + \alpha_{pn}(k_n-l_n).
\end{equation}
Indeed, one sees that the entire expression factors,
so that
\begin{equation}
A_{\alpha} = A_{\alpha_1} \otimes \cdots \otimes A_{\alpha_n},
\end{equation}
where $\alpha_j = (\alpha_{qj},\alpha_{pj})$ and
$A_{\alpha_j}$ is the usual phase-point operator
of Eq.~(\ref{Adefodd}) for an $r \times r$ phase space.
(See the related analyses in Refs.~\cite{Klimov,Vourdas4,Pittenger}.)
In this form, the phase-point operators look like those
of Ref.~\cite{Wootters}, where the phase space is a 
$2n$-dimensional vector space over ${\mathbb Z}_r$.
However, there is a conceptual difference between
the two formalisms, concerning
the way the phase space is partitioned to define marginal probabilities.
In the higher-dimensional phase space, it is natural to
consider an $n$-dimensional ``slice,'' which is an ordered
$n$-tuple of lines from the $r \times r$ phase
spaces of the individual particles. The sum of the
phase-point operators over such a slice is the 
projection operator onto a pure product state.
In contrast, the sum of the $A_\alpha$'s over 
a {\em line} in the two-dimensional phase space 
over ${\mathbb F}_d$ is the projection operator onto
a pure state that may well be entangled.  As in the
prime-dimensional case, each of the $d+1$ striations defines
an orthogonal basis for the Hilbert space, and these
$d+1$ bases are mutually unbiased.  

Once we adopt this two-dimensional phase space, with
the phase-point operators given by Eq.~(\ref{Adeffield}),
one can check that every step of our arguments
in Sections \ref{quasitoactual} and \ref{mainsection}
goes through exactly as before.  In particular, we
use the fact that, as in the odd prime case, every
symplectic matrix $S$ is {\em legal} in the sense that there
exists a unitary transformation $U_S$ such that \cite{Appleby}
\begin{equation}
U_S A_\alpha U_S^\dag = A_{S\alpha}.
\end{equation}
The displacement operators, defined 
by
\begin{equation}
D_\alpha = D_{\alpha_1} \otimes \cdots \otimes D_{\alpha_n},
\end{equation}
still satisfy Eq.~(\ref{displacementproperty}).  It is 
simply that the symbols representing phase-space 
points are now elements of ${\mathbb F}_d^2$. 

Thus, for an experiment consisting of a preparation,
channel, and measurement, we can still take as our 
framework the triple ${\mathcal F}=(B',S,B)$, where $B$ and $B'$
are striations and $S$ is a symplectic matrix.  
The classical experiment is still described by
$R^B(\alpha | {w})$, $R^S_{\mathcal E}(\beta | \alpha)$,
and $R^{B'}(E | \beta)$, all given by the same
formulas as before.  And the actual quantum prediction
for the probability of the outcome $E$ is still given
by
\begin{equation}
\Delta P(E| \mathcal{E},{w})
= \frac{1}{\mathcal Z} \sum_{\mathcal F}
\Delta R^{\mathcal F}(E | \mathcal{E},{w}),
\end{equation}
with ${\mathcal F}= (B',S,B)$ and the classical prediction given by
\begin{equation}
R^{\mathcal F}(E | \mathcal{E},{w})
= \sum_{\beta,\alpha}
R^{B'}(E|\beta)R^S_{\mathcal E}(\beta | \alpha)
R^B(\alpha | {w}).
\end{equation}
The redundancy factor ${\mathcal Z}$ is again equal
to $d$ if we use the full symplectic group.  If there
exists a smaller ``minimal reconstructing set,'' then
we could use that set and have ${\mathcal Z}=1$.  We
are not aware of work either proving or 
disproving the existence of such a special set
of symplectic matrices for any value of
$d$ equal to $r^n$ with
$r$ an odd prime and $n > 1$.  

Suppose we have the complete description of a preparation
for a system of $n$ particles, each with odd-prime Hilbert-space
dimension $r$.  That is, we have all the probability
distributions $R^B(\alpha | {w})$.  Is there a simple
way to obtain from this description a similar description
of the effective preparation of just one of the
particles, say, the first one?  In Hilbert space
language, we would take a partial trace of the 
whole system's density matrix.  What does this
partial trace look like in our formalism?

We begin by addressing this question at the level
of the Wigner function.  In terms of the Wigner function,
the density matrix of the
whole system is given by
\begin{equation}
\begin{split}
{w} &= \sum_\alpha Q(\alpha|{w})A_\alpha \\
&=\sum_{\alpha}Q(\alpha|{w})
A_{\alpha_1} \otimes \cdots \otimes A_{\alpha_n}.
\end{split}
\end{equation}
Taking the partial trace of ${w}$ to get
the state ${w}_1$ of the first particle,
we have
\begin{equation}
{w}_1 = \hbox{tr}_{2,\ldots,n}{w}
= \sum_{\alpha_1}\left[\sum_{\alpha_2,\ldots,\alpha_n}Q(\alpha_1,\ldots,\alpha_n|{w})\right] A_{\alpha_1}.
\end{equation}
Evidently the quantity in square brackets is playing the
role of the Wigner function for the single-particle
state ${w}_1$.  That is, we can write
\begin{equation}
Q(\alpha_1|{w}_1) = \sum_{\alpha_2,\ldots,\alpha_n}Q(\alpha_1,\ldots,\alpha_n|{w}).
\end{equation}
Thus as far as the Wigner function is concerned, we
take the partial trace by summing over the 
phase-space coordinates of the traced-out particles.  
Since $\alpha_1$ is a vector in ${\mathbb F}_r^2$,
it is perfectly appropriate for labeling the phase-space
points of particle 1. (Note: We are not making a similar claim
for {\em all} examples of partial tracing.  If we
trace over all but the first two particles, for example,
we are left with $\alpha_1$ and $\alpha_2$, but the
expression $\alpha_{q1} e_1 + \alpha_{q2} e_2$ does not 
make sense in the field ${\mathbb F}_{r^2}$,
because $e_1$ and $e_2$ are not elements of that
field.)

We now translate the Wigner-function picture into
the language of the $R^B$'s.  We begin with a combination
of Eqs.~(\ref{PBfromQ}) and (\ref{RB-eq}):
\begin{equation} \label{QtoRB}
R^B(\alpha|{w}) = \frac{1}{d}\sum_{\beta \in \ell}Q(\beta|w),
\end{equation}
where $\ell$ is the line in $B$ that contains the point
$\alpha$.
Let us define $R^B(\alpha_1 | {w}_1)$ by summing
over $\alpha_2, \ldots, \alpha_n$.
\begin{equation}
R^B(\alpha_1 |{w}_1) = \sum_{\alpha_2,\ldots,\alpha_n}
R^B(\alpha | {w}).
\end{equation}
The notation indicating that this expression depends 
only on ${w}_1$ (and not on any other aspects of 
the full density matrix ${w}$)
is justified by the relation
(\ref{QtoRB}) and the fact that $Q(\alpha_1|{w}_1)$
depends only on ${w}_1$. Now we use Eq.~(\ref{DeltaDelta})
(which is still valid in this new context)
to write
\begin{equation}
\begin{split}
Q(\alpha_1|{w}_1) &=\sum_{\alpha_2,\ldots,\alpha_n}
Q(\alpha_1,\ldots,\alpha_n|{w}) \\
&= \sum_{\alpha_2,\ldots,\alpha_n} \left[\sum_B R^B(\alpha_1,\ldots,\alpha_n|{w}) - \frac{1}{d}\right]\\
&=\sum_B R^B(\alpha_1|{w}_1) - \frac{r^{2(n-1)}}{r^n}.
\end{split}
\end{equation}
With the substitutions
\begin{equation}
\begin{split}
Q(\alpha_1|{w}_1) &= \Delta Q(\alpha_1|{w}_1) + \frac{1}{r^2}, \\
R^B(\alpha_1|{w}_1) &= \Delta R^B(\alpha_1|{w}_1) + \frac{1}{r^2},
\end{split}
\end{equation}
we find that all the additive constants drop out,
and we are left with
\begin{equation} \label{toomanyBs}
\Delta Q(\alpha_1|{w}_1) = \sum_B \Delta R^B(\alpha_1|{w}_1).
\end{equation}

Now, Eq.~(\ref{toomanyBs}) is {\em not} in our standard
form for an $r$-dimensional particle, because the
superscript $B$ still ranges over all the striations
of the large phase space ${\mathbb F}_d^2$.  However,
we can convert this expression to our standard form
via the following facts, which we prove in 
Appendix C.  (i) If the slope $m$ of the striation
$B$ is of the form $m = m_1 e_1^{-1}\tilde{e}_1$ with
$m_1 \in {\mathbb Z}_r$, or if the slope is infinite,
then 
$R^B(\alpha_1 | {w}_1) = R^{B_1}(\alpha_1 | {w}_1)$,
where $B_1$ is the striation in ${\mathbb Z}_r^2$ whose 
slope is $m_1$, or infinity if $m=\infty$.  (ii) If the slope $m$ is of the form
$m = \sum_j m_j e_1^{-1}\tilde{e}_j$
with $m_j \in {\mathbb Z}_r$ and $m_j\ne 0$ for
some $j\ne 1$, then 
$\Delta R^B(\alpha_1 | {w}_1) = 0$, so those
striations do not contribute to the sum in 
Eq.~(\ref{toomanyBs}).  This leaves us with
\begin{equation} \label{singleparticleQR}
\Delta Q(\alpha_1|{w}_1) = \sum_{B_1} \Delta R^{B_1}(\alpha_1|{w}_1),
\end{equation}
which is in our standard form.
In this way, we can go from a description of the
$n$-particle system to a description of the
state of the first particle.  That is, we
have a way of ``tracing out'' the other 
$n-1$ particles, and the resulting picture accords
with our earlier treatment of a particle with
a prime-dimensional Hilbert space.  

We illustrate these ideas by working out a specific
example in detail.  Consider a pair of qutrits
in the pure, entangled state
\begin{equation} \label{psi}
|\psi\rangle = \frac{1}{2\sqrt{2}}(|00\rangle
+ |01\rangle + |10\rangle + |11\rangle)
+ \frac{1}{\sqrt{2}}|22\rangle.
\end{equation}
Its Wigner function will be a real function on
the $9 \times 9$ phase space ${\mathbb F}_9^2$.
We construct the field ${\mathbb F}_9$ by
identifying a quadratic polynomial, with 
coefficients in ${\mathbb F}_3$, that has no
roots in ${\mathbb F}_3$.  Such a polynomial
is $x^2 + 1$.  Let us then define $\xi$ to be such
that $\xi^2 + 1 = 0$; that is, $\xi^2 = 2$.  
(This is analogous to defining the imaginary
unit $i$ to satisfy $i^2 + 1 = 0$, which no
real number satisfies.)  A general
element of ${\mathbb F}_9$ is then of the form
$a + b\xi$ with $a,b \in {\mathbb F}_3$.

In order to treat our $d=9$ system as two
distinct qutrits, we choose a field basis in which
to write the horizontal phase-space coordinate.  
Let us choose the basis $(e_1,e_2) = (1, \xi)$.  One finds that
the dual basis is $(\tilde{e}_1,\tilde{e}_2)
= (2, \xi)$, and we use this dual basis to expand
the vertical phase-space coordinate.  With these
conventions, we show here the Wigner function of the
state $|\psi\rangle$ given in Eq.~(\ref{psi}).

\begin{equation}
\begin{split}
&\setlength{\tabcolsep}{2.1pt}
\renewcommand{\arraystretch}{1.13}
\begin{tabular}{cccc}\begin{tabular}{r}\scriptsize{$1+2\xi$} \\  \scriptsize{$1+\xi$} \\   \scriptsize{$1$} \\   \scriptsize{$2 + 2\xi$} \\   \scriptsize{$2 + \xi$} \\   \scriptsize{$2$}  \\   \scriptsize{$2\xi$} \\   \scriptsize{$\xi$}  \\   \scriptsize{$0$} \end{tabular} & \begin{tabular}{c}22\\  21\\  20\\  12\\  11\\  10\\  02\\ 01\\  00 \end{tabular} &
\setlength{\tabcolsep}{4pt}
\renewcommand{\arraystretch}{1.08}\begin{tabular}{!{\color{black}\vrule width 1pt}c|c|c!{\color{black}\vrule width 1pt}c|c|c!{\color{black}\vrule width 1pt}c|c|c!{\color{black}\vrule width 1pt}}
\Xhline{2.25\arrayrulewidth}
\arrayrulecolor{black} -1 & 5 & -1 & 5 & -1 & -1 & -1& -1 & 5 \\ 
\hline 5 & -1 & -1 & -1 & 5 & -1 & -1 & -1 & 5 \\ 
\hline -1 & -1 & 2 & -1 & -1 & 2 & -1 & -1 & 2 \\
\Xhline{2.25\arrayrulewidth}
5 & -1 & -1 & -1 & 5 & -1 & -1 & -1 & 5 \\
\hline -1 & 5 & -1 & 5 & -1 & -1 & -1 & -1 & 5 \\
\hline -1 & -1 & 2 & -1 & -1 & 2 & -1 & -1 & 2 \\
\Xhline{2.25\arrayrulewidth} -1 & -1 & -1 & -1 & -1 & -1 & 2 & 2 & 2 \\
\hline -1 & -1 & -1 & -1 & -1 & -1 & 2 & 2 & 2 \\
\hline 5 & 5 & 2 & 5 & 5 & 2 & 2 & 2 & \hspace{0.5mm}8\hspace{0.5mm}  \\
\Xhline{2.25\arrayrulewidth}
\end{tabular} & $\times \frac{1}{72}$ \end{tabular} 
\\
&\setlength{\tabcolsep}{3.45pt}
\hspace{1.9cm} \begin{tabular}{ccccccccc}00  & 01 &  02 & 10 & 11 & 12 & 20 & 21 & 22 \end{tabular}
\\
&\setlength{\tabcolsep}{1.3pt}
\hspace{1.85cm} \begin{tabular}{ccccccccc}  \hspace{1mm}  \scriptsize{0}\hspace{1mm}  & \hspace{1mm}  \scriptsize{$\xi$}  \hspace{1mm} &  \hspace{0mm}  \scriptsize{$2\xi$}  \hspace{0mm} & \hspace{1mm}  \scriptsize{1} \hspace{1mm}  & \scriptsize{$1\hspace{-1mm}+\hspace{-1mm}\xi$} & \scriptsize{$1\hspace{-1mm}+\hspace{-1mm}2\xi$}  & \hspace{1mm}  \scriptsize{2} \hspace{1mm}  & \scriptsize{$2\hspace{-1mm}+\hspace{-1mm}\xi$}  & \scriptsize{$2\hspace{-1mm}+\hspace{-1mm}2\xi$}  \end{tabular}
\end{split}
\end{equation}
The first line of labels along the horizontal
axis shows the pairs $(\alpha_{q1},\alpha_{q2})$.  The 
line below it shows the field values $\alpha_q
= \alpha_{q1}e_1 + \alpha_{q2}e_2$.  Similarly, along
the vertical axis we have $(\alpha_{p1},\alpha_{p2})$
and $\alpha_p = \alpha_{p1}\tilde{e}_1 + \alpha_{p2}\tilde{e}_2$.

From the Wigner function, we can construct the
probability distribution $R^B(\alpha | {w})$ for
any striation $B$.  Here we choose the striation
with slope $m=2$.  So we sum the Wigner function
over each line with slope 2, and we spread this
sum uniformly over the line to get 
$R^B(\alpha | {w})$.  Here is the result:
\begin{equation}
\begin{split}
&\setlength{\tabcolsep}{2.1pt}
\renewcommand{\arraystretch}{1.13}
\begin{tabular}{cccc}\begin{tabular}{r}\scriptsize{$1+2\xi$} \\  \scriptsize{$1+\xi$} \\   \scriptsize{$1$} \\   \scriptsize{$2 + 2\xi$} \\   \scriptsize{$2 + \xi$} \\   \scriptsize{$2$}  \\   \scriptsize{$2\xi$} \\   \scriptsize{$\xi$}  \\   \scriptsize{$0$} \end{tabular} & \begin{tabular}{c}22\\  21\\  20\\  12\\  11\\  10\\  02\\ 01\\  00 \end{tabular} &
\setlength{\tabcolsep}{5.5pt}
\renewcommand{\arraystretch}{1.08}\begin{tabular}{!{\color{black}\vrule width 1pt}c|c|c!{\color{black}\vrule width 1pt}c|c|c!{\color{black}\vrule width 1pt}c|c|c!{\color{black}\vrule width 1pt}}
\Xhline{2.25\arrayrulewidth}
\arrayrulecolor{black} 0 & 1 & 1 & 4 & 0 & 0 & 0 & 1 & 1 \\ 
\hline 1 & 0 & 1 & 0 & 4 & 0 & 1 & 0 & 1 \\ 
\hline 1 & 1 & 0 & 0 & 0 & 4 & 1 & 1 & 0 \\
\Xhline{2.25\arrayrulewidth}
4 & 0 & 0 & 0 & 1 & 1 & 0 & 1 & 1 \\
\hline 0 & 4 & 0 & 1 & 0 & 1 & 1 & 0 & 1 \\
\hline 0 & 0 & 4 & 1 & 1 & 0 & 1 & 1 & 0 \\
\Xhline{2.25\arrayrulewidth} 0 & 1 & 1 & 0 & 1 & 1 & 4 & 0 & 0 \\
\hline 1 & 0 & 1 & 1 & 0 & 1 & 0 & 4 & 0 \\
\hline 1 & 1 & 0 & 1 & 1 & 0 & 0 & 0 & 4  \\
\Xhline{2.25\arrayrulewidth}
\end{tabular} & $\times \frac{1}{72}$ \end{tabular}
\\
&\setlength{\tabcolsep}{3.35pt}
\hspace{1.9cm} \begin{tabular}{ccccccccc}00  & 01 &  02 & 10 & 11 & 12 & 20 & 21 & 22 \end{tabular}
\\
&\setlength{\tabcolsep}{1.2pt}
\hspace{1.85cm} \begin{tabular}{ccccccccc}  \hspace{1mm}  \scriptsize{0}\hspace{1mm}  & \hspace{1mm}  \scriptsize{$\xi$}  \hspace{1mm} &  \hspace{0mm}  \scriptsize{$2\xi$}  \hspace{0mm} & \hspace{1mm}  \scriptsize{1} \hspace{1mm}  & \scriptsize{$1\hspace{-1mm}+\hspace{-1mm}\xi$} & \scriptsize{$1\hspace{-1mm}+\hspace{-1mm}2\xi$}  & \hspace{1mm}  \scriptsize{2} \hspace{1mm}  & \scriptsize{$2\hspace{-1mm}+\hspace{-1mm}\xi$}  & \scriptsize{$2\hspace{-1mm}+\hspace{-1mm}2\xi$}  \end{tabular}
\end{split}
\end{equation}

Now, is this striation, with slope $m=2$, a striation
that contributes to Eq.~(\ref{singleparticleQR})? The
test is to expand $m$ in the basis $(e_1^{-1}\tilde{e}_1,
e_1^{-1}\tilde{e}_2)$---that is, to find $m_1$
and $m_2$ in $m = m_1 e_1^{-1}\tilde{e}_1 + m_2 e_1^{-1}\tilde{e}_2$---and to see whether $m_2$ is zero.
If it is, then this striation does contribute to 
Eq.~(\ref{singleparticleQR}), and the corresponding
striation in the small phase space ${\mathbb F}_3^2$
is the one with slope $m_1$.  Upon doing the expanding,
we get $m = 2 = 1(2) + 0(\xi)$.  So indeed, this striation
does contribute, and the corresponding striation has
slope $m_1 = 1$.  

We can see how this works out by summing
$R^B(\alpha | {w})$ over each of the $3 \times 3$
grids bounded by thick lines.  The result is
\begin{equation}
\begin{split}
&\setlength{\tabcolsep}{1.9pt}
\begin{tabular}{ccc}
\begin{tabular}{c}
2 \\ 1 \\ 0 \end{tabular}
\setlength{\tabcolsep}{4pt}
\begin{tabular}{|c|c|c|}
\hline
1 & 2 & 1 \\
\hline
2 & 1 & 1 \\
\hline
1 & 1 & 2 \\
\hline
\end{tabular} & $\times \frac{1}{12}$
\end{tabular} \\
&\hspace{6mm}\setlength{\tabcolsep}{4.3pt}\begin{tabular}{ccc}
0 & 1 & 2 
\end{tabular} \hfill
\end{split}
\end{equation}
One can check that this is indeed the correct
distribution $R^{B_1}(\alpha_1 | {w}_1)$ for the
reduced density matrix ${w}_1 = \hbox{tr}_2 {w}$
and for the striation $B_1$ with slope $m_1 = 1$. 

For the above construction, it is not necessary
that the dimension of each individual particle
be prime.  Suppose we have a system of $n$ particles, each
of which has Hilbert-space dimension $r$, where
$r$ is a nontrivial {\em power} of an odd prime $p$.  So $d=r^n$ with $r=p^s$.  
The phase space for a single particle will be
${\mathbb F}_r^2$.  To get the phase space for
the whole system, we form the extension field
${\mathbb F}_{r^n}$.  This field is none other
than ${\mathbb F}_d$---there
is only one field with $d$ elements---but
we wish to think of the field as an extension of
${\mathbb F}_r$.  To obtain phase-space coordinates
for the individual particles, we expand the horizontal
coordinate $\alpha_q$ in a basis $(e_1, \ldots, e_n)$.
That is, $\alpha_q = \sum_j \alpha_{qj}e_j$, where 
$\alpha_{qj}$ is an element of ${\mathbb F}_r$.  We 
expand the vertical coordinate $\alpha_p$ in the 
dual basis $(\tilde{e}_1, \ldots, \tilde{e}_n)$.
Here the field trace used to define the dual is
$\hbox{Tr}_{\mathbb{F}_d/\mathbb{F}_r}$,
a function from ${\mathbb F}_{d}$ to ${\mathbb F}_r$
(and not a function from ${\mathbb F}_{d}$ to
${\mathbb F}_p$).
It can be defined either by Eq.~(\ref{trace2})
or by Eq.~(\ref{trace1}).
We ultimately take the absolute trace, taking values in
${\mathbb F}_p$, in the definition of 
the phase-point operators, which we now write 
as
\begin{equation}
(A_\alpha)_{kl} = \delta_{2\alpha_q, k+l}\, \omega^{\hbox{\scriptsize Tr}_{\mathbb{F}_d/\mathbb{F}_p}[\alpha_p (k-l)]}
\end{equation}
with $\omega$ equal to $e^{2\pi i/p}$.
With these definitions,
one finds that 
everything we have done for odd prime values
of $r$ continues to work.  In particular, starting
with a state of $n$ particles as represented by
a collection of $R^B$ distributions, we can take appropriate
sums over the coordinates of the last $n-1$ particles
to get the state of the first particle.  That is, our
method of taking the partial trace remains valid in this
more general setting.

\section{Challenges}  \label{challenges}

The formalism we have developed cannot be 
immediately extended to all values of the 
Hilbert-space dimension $d$.  In this paper we have relied
heavily, though often implicitly, on the properties of finite fields, and 
when $d$ is not a prime power, there exists no
finite field having exactly $d$ elements.  
In the present section, we describe some of the
difficulties one encounters for values of $d$ other
than those we have considered. 

{We begin with a bit of parameter
counting.  A general unital channel requires
$(d^2 - 1)^2$ parameters for its specification.
For a fixed symplectic matrix $S$, the probability distribution $P^S_{\mathcal E}(\delta)$
provides $d^2 - 1$ parameters, since there are $d^2$
values of $\delta$ but the probabilities must sum
to unity.  So we need at least $d^2 - 1$ legal
symplectic matrices in order for the whole set
of $P^S$'s to have a chance of representing a general unital channel.  A minimal reconstructing set comprises
precisely this many matrices.}

Now consider a composite dimension that is not
a prime power,
such as $d=15=3\cdot 5$.  {By using as our phase space
the Cartesian product of ${\mathbb Z}_3^2$ and
${\mathbb Z}_5^2$, we can in fact generalize our treatment
of preparations and measurements to this case,
for example representing a preparation $w$ by
probability distributions $R^{B_1,B_2}(\alpha_1,\alpha_2 | w)$, where the $B$'s and $\alpha$'s refer to
striations and points in
${\mathbb Z}_3^2$ and
${\mathbb Z}_5^2$. (We plan to present the details
of this generalization in a future paper.)}
Our main difficulty arises
in the description of a channel.  It is perfectly 
sensible to speak of the group of symplectic 
matrices with entries in ${\mathbb Z}_{15}$ \cite{Gross1}.
{This group factorizes into the symplectic
groups represented by matrices with entries in
${\mathbb Z}_3$ and ${\mathbb Z}_5$.  
If we try to form a minimal reconstructing set,
starting with the minimal reconstructing sets
for dimensions 3 and 5, we end up with only
$(3^2 - 1)(5^2 - 1)$ matrices, fewer than the
$(15^2 - 1)$ that we need.  We could try using
the full symplectic group, which has more than
enough matrices, but even then, one can show that
it is impossible to reconstruct $Q_{\mathcal E}(\beta | \alpha)$
from the whole set of distributions 
$P^{S_1,S_2}_{\mathcal E}(\delta_1,\delta_2)$.}
This fact does not imply that there
is no extension of our work to the case $d=15$, only
that there is no obvious, immediate extension.  

Dimensions that are powers of 2, as when the system
is a collection of qubits, also present a challenge,
even though there does exist a field of order $2^n$ 
for every positive integer $n$.  In this case, the
challenge comes from the scarcity of legal
symplectic matrices.  We have already seen that
for $d=2$, only three of the six symplectic matrices
are legal in the sense that they correspond to 
unitary transformations of the phase-point operators
(Eq.~(\ref{SUcorrespondence})).  This is not a problem for $d=2$,
because three is precisely the number of symplectic
matrices we need.  However, it is already a problem
for $d=4$, e.g., for a two-qubit system.

In dimension 4, it is certainly possible to use as
a phase space the two-dimensional vector space
over the field ${\mathbb F}_4$, in analogy with
what we have done in Section \ref{primepower}.  
However, in contrast to the case of odd-prime-power
dimension, there is
no single, natural choice of phase-point operators for 
dimension 4.  Within the framework of Ref.~\cite{Gibbons},
there are, for $d=4$, four truly distinct ways of defining
the phase-point operators (four similarity classes,
in the terminology of Ref.~\cite{Gibbons}).  
There are exactly 60 symplectic $2 \times 2$ 
matrices with entries in ${\mathbb F}_4$, and for
any given definition of the phase-point operators,
one
can check to see how many of these 
symplectic matrices are legal.
One finds that for two of the four possible definitions,
there are only three legal matrices, and for the other
two, there are only five legal matrices.  But we 
need $4^2 - 1 = 15$ legal matrices $S$ to characterize
a general unital channel.

%Alternatively, we could treat the case $d=4$ by using
%a four-dimensional phase space over the field
%${\mathbb Z}_2$.  The phase-point operators would be
%defined as the tensor products of the usual single-qubit
%phase-point operators.  Now, there are
%720 symplectic $4 \times 4$ matrices, but we find
%that only 18 of them are legal.  (They are all 
%possible combinations of the 3 legal symplectics
%on qubit \#1, the 3 legal symplectics on qubit
%\#2, and the swap operation.)  One can be pleased
%that 18 is not smaller than 15, but it is still the
%case that the 18 values of $S\alpha$, for any fixed
%nonzero $\alpha$, are not able to reach all the nonzero %points. So again, our methods cannot be immediately
%applied to the case of two qubits.  

There are various avenues one could follow toward
generalizing our formalism.  One possibility, which
might help in trying to build a formalism for $d=15$
out of the cases $d=3$ and $d=5$, is to
extend the list of classical frameworks in our 
description of a channel.  In Fig.~2, one can discern in the $4 \times 4$ grids
a set of three striation-like
partitions, for which each ``line'' of one
partition intersects each line of any other partition
in exactly one point.  It is possible to add two other
partitions of the grid and still retain this relation
between any pair of partitions.  The new ones would 
consist of
the horizontal and vertical lines.  
Should we look for a way of including these other two partitions
in our picture?  (For this single-qubit case, the partitions can in fact be identified with striations of $\mathbb{F}_4^2$.) Alternatively, and especially 
for dealing with even values of $d$,
one could try switching from a $d \times d$ phase space
to a $2d \times 2d$ phase space \cite{Leonhardt1,Leonhardt2}. This phase space and the associated Wigner function were designed specifically to accommodate even values of the
Hilbert-space dimension. 

A less elegant solution is simply to note that 
every system with a finite-dimensional Hilbert space
can be accommodated within any Hilbert space of
higher dimension.  So, if we want to treat within
our formalism a system
of dimension 6, we treat it as a system with dimension 7.
In this sense, what we have already done is sufficient
for the quantum description of any discrete system. 

The case of continuous phase space, applicable to
a system with an infinite-dimensional Hilbert space,
is clearly a case that needs to be considered.  
Much work has been done on the
representation of quantum states in terms of the 
marginal distributions over such a phase space \cite{Ibort}.
The generalization of our treatment of {\em channels}
to continuous phase space will likely raise new questions.
We plan to treat this case in future work.

\section{Conclusions}  \label{lastsection}

We begin this section by writing down our partial 
formulation of quantum theory in a self-contained way,
without reference to Hilbert space or quasiprobabilities.

First, rather than characterizing a discrete system by
its Hilbert-space dimension, we adopt an operational
definition of $d$: it is the maximum number of outcomes
of a repeatable measurement, that is, a measurement
that yields the same outcome when repeated.  In what follows, we 
consider only systems with odd-prime-power values 
of $d$.  
Also, we will now interpret the symbols ${w}$,
$\mathcal{E}$ and $E$ purely operationally in the sense that ${w}$ refers solely to a preparation procedure, $\mathcal{E}$ to
a physical channel, and $E$ to a measurement outcome.  None
of these symbols refers to an operator on Hilbert space or
an operation on Hilbert-space operators. 
In these terms, we express the basics of
operational quantum theory as follows.  

A preparation $w$ is represented by a collection of
probability distributions $R^B(\alpha|{w})$, where $\alpha$
takes values in the $d \times d$ discrete phase space
${\mathbb F}_d^2$ and $B$ is a striation of that
space.  
% (We now avoid using the symbol
% ${w}$ for the preparation, since it strongly suggests
% an operator on Hilbert space.  We could use a more
% neutral symbol, say, $\Pi$, but we choose instead to leave the
% specification of the preparation implicit to
% make the notation less cluttered.)  
Each distribution
$R^B$ must satisfy an epistemic constraint: it must be uniform over each line of the 
striation $B$.   

A channel $\mathcal{E}$ is represented by a collection of conditional
probability distributions $R^S_{\mathcal E}(\beta | \alpha)$, where
$S$ is a symplectic matrix. 
Each such distribution must satisfy an epistemic
constraint: it can depend only on the displacement
$\delta = \beta - S\alpha$.  

A measurement outcome $E$ is represented by a collection 
of conditional
probability functions $R^{B'}(E|\beta)$.  Each such
function must satisfy an epistemic constraint: it
must be uniform over each line in the striation $B'$.  

In addition to the epistemic constraints, the
$R$ functions describing each component of the 
experiment must also satisfy certain
global constraints, as we have noted in the Introduction.  
For example, the distributions
$R^B(\alpha|{w})$ cannot be chosen independently of each
other.  These additional constraints are specified
in Appendix D.  

Once all the $R$ functions have been given, we compute,
for each {\em coherent} framework, the classical probability
of the outcome $E$.  We do this using the ordinary
rules of probability theory.  If the experiment
includes a single channel, this probability is
\begin{equation}  \label{normalprobs}
R^{\mathcal F}(E|{w}) = \sum_{\beta,\alpha} R^{SB}(E|\beta)R^S(\beta|\alpha)R^B(\alpha|{w}).
\end{equation}
We then compute the {\em quantum} probability of the outcome
$E$ via the formula
\begin{equation}  \label{weirdrule}
\Delta P(E|{w}) = \frac{1}{\mathcal Z} \sum_{\mathcal F}\Delta R^{\mathcal F}(E|{w}).
\end{equation}

Though this description is self-contained, we again note 
that the 
constraints written down in Appendix D are not as
simple as we would like.  They are based on 
the analogous constraints on the Wigner function 
and the transition quasiprobabilities, and they appear simpler in that quasiprobabilistic
setting than in our current setting.  
{In a separate paper, we explore the idea
that the global constraints may follow
%that our whole construction, including the constraints, may flow 
from a short list of principles {\cite{Braasch2022}}. 
Our central principle can be stated informally in this way: 
to find the probability functions representing 
the combination of two components of an experiment, sum the nonrandom parts of the classically expected results over every relevant framework.
We show that when this idea is augmented with a few modest assumptions, we can at least reconstruct the
quantum theory of a single qubit.  
%end up with the physical description of a qubit for the case of $d=2$.
We hope that these or similar principles prove to be sufficient to single out quantum theory when $d$ is a prime greater than 2.}
% It is worth noting that one of the assumptions we adopt relates to how one must choose the set of allowable preparations in our model.
% This }
% We hope it is
% possible to find an alternative set of constraints, 
% more native, as it were, to the $R$ description, that
% are equivalent to the constraints in Appendix D.  

It is worth noting that with a complete collection of classical descriptions of an experiment,
%experimental steps (i.e. the preparations, channels, and measurements), 
the description of the quantum experiment is always complete at every step. 
%In other words, one can move between %the description of a quantum experiment using
%the collection of classical experiments and the standard Hilbert space theory at any point freely. 
For example, at any point
as we move horizontally through Fig.~3, we can translate
the restricted classical descriptions into a quantum 
description, and it will be the correct quantum 
description of that phase of the experiment.  Indeed,
this fact is the basis of the central principle
described in the preceding paragraph.

{It is interesting also to think
about how
our formalism applies to 
quantum computation.
We have seen how our construction can describe
a quantum system with Hilbert space dimension
equal to an odd prime power.  A quantum computer
consisting of $d$-dimensional units, with prime $d$, is such a 
system, so we can imagine a quantum computation
implemented as a series of gates acting on 
a standard initial state of this system.  Within a given framework, we could
access the output statistics of one ``branch'' of
the quantum computation by sampling from the 
initial state (a probability distribution 
$R^B(\alpha|w)$ over
phase space) and carrying out the stochastic
process defined by the transition probabilities
$R^{S_j}_{\mathcal{E}_j}(\beta | \alpha)$ for the series of gates.
However, to capture the full quantum computation,
we would need to keep track of all the frameworks,
and the number of frameworks grows exponentially
both with the number of qudits and with the number
of gates.}

{
One can ask what error is
incurred if we try to get away with considering
only a subset of the frameworks and extrapolating
from those results.  It can certainly be the
case that some frameworks are more important
than others; for example, if one of our gates is the 
unitary transformation $U_S$ associated with
a symplectic transformation $S$, then
the framework indexed by that $S$ is the only
one we need to consider for that gate.  So there
may be a clever way of limiting the number of
frameworks.
Our unusual rule for combining probabilities
raises additional questions.  For example, if we
have the ability
to sample from each of a set of probability
distributions, can we efficiently simulate
the effect of sampling from the distribution
obtained by summing the nonrandom parts?
And is there a sense in which this summation
procedure substitutes for 
the negativity that has been
identified as the key to the power of quantum 
computation \cite{Veitch2012, Delfosse2015, Raussendorf2017}? 
We leave these questions
for future investigations.}

As we mentioned in the Introduction, we do not claim
to know how to extract from our work an answer to the 
question: what is really going on?  A simple classical
theory with an epistemic constraint {\em can} be read
as answering that question: what is really going on 
is that the system is always in one of the ontic states,
but a measurement disturbs the system, and the
epistemic constraint prevents the observer from ever
knowing the system's true state.  For the description
we are adopting here, we have no such account of
reality to offer.  Each of our imagined {\em classical} observers
might construct for themselves such an account---their
computed probability $R^{\mathcal F}(E|{w})$ can be perfectly well
explained by a picture of the system jumping around
in phase space---but
we, who are assembling the predictions of
{\em all} these classical observers, cannot 
explain the quantum probability $P(E|{w})$ in this way.
If we were forced to put our prediction algorithm
into the standard probabilistic form (as in 
Eq.~(\ref{normalprobs}), for example), we would 
naturally be led 
to use {\em quasiprobabilities}; that is, we
would return to the Wigner-function formulation
of quantum theory.  For us, the cost
of a description based on ordinary probability 
distributions, as opposed to quasiprobabilities,
is that we must use the rule (\ref{weirdrule}),
which, though mathematically simple and
conceptually intriguing, is
difficult to interpret as a way of combining
probabilities.  Indeed, if it were not for the
constraints on the relationships among the
$R$ distributions (Appendix D), this rule for
combining probabilities could easily yield
a probability greater than 1 or less than 0.

{This is not to say that no
ontological interpretation of our formalism can
be found.  Our picture lies outside the standard
framework of ontological models \cite{Harrigan2010},
but generalizations of that framework have
been proposed \cite{Leifer}
% ---one notable possibility
% is to allow retrocausation \cite{Argaman1, Argaman2}---
and it is
conceivable that one can construct a 
picture of reality that gives physical meaning
to the mathematics we have presented here.  
As we have suggested above, we can reasonably hope to 
find a natural way of explaining the global
constraints listed in Appendix D.  What remains, then,
is to find a physical interpretation of the
enigmatic formula (\ref{weirdrule}) by which
the predictions from our many classical
worlds are to be combined.}

%We are by no means arguing that our work rules 
%out the possibility of an epistemic understanding of 
%the quantum state.  (Others have addressed that
%question---see, for example, the review
%article \cite{Leifer}.)  It is only that, if we want
%to base quantum theory on the particular
%epistemically restricted theory we are considering here,
%the ``quantumness'' of the theory still 
%manages to find a way of expressing itself: it
%makes itself known in
%the enigmatic formula (\ref{weirdrule})
%and in the global constraints on the 
%classical probability distributions.
%It is this formula that plays the role of
%the classical-to-quantum
%bridge in our story.  

\appendix

\section{Proof of a Kronecker-delta equality}
Here we prove Eq.~(\ref{factforappendixA}), which we copy here:
\begin{equation}  \label{factforappendixAagain}
\frac{1}{\mathcal{Z}d^2}\sum_{S,\delta}
\delta_{\beta,S\alpha+\delta}\delta_{\beta',S\alpha'+\delta}
= \delta_{\beta\beta'}\delta_{\alpha\alpha'}
+ \frac{1}{d^2}(1-\delta_{\beta\beta'} 
- \delta_{\alpha\alpha'} ).
\end{equation}
We start with the left-hand side (which we call ``LHS'' in what follows) and use one
of the Kronecker deltas to do the sum over $\delta$. This gives us
\begin{equation}  \label{Ssum}
\hbox{LHS} =
\frac{1}{{\mathcal Z}d^2} \sum_S \delta_{S(\alpha' - \alpha), (\beta' - \beta)}.
\end{equation}
We now use a nice property of the symplectic group
$Sp(2,\mathbb{F}_d)$ where $\mathbb{F}_d$ is any finite field,
such as $\mathbb{Z}_d$ when $d$ is prime \cite{Chau}: for any
nonzero points $\mu$ and $\nu$ in phase space,
\begin{equation} \label{sumtod}
\sum_S \delta_{S\mu, \nu} = d.
\end{equation}
That is, as $S$ ranges over all symplectic matrices,
$S\mu$ hits every nonzero point exactly $d$ times.
The result is even simpler if, instead of using the
full set of symplectic matrices, we use a 
``minimal reconstructing set,'' that is, a set of just
$d^2 - 1$ symplectic matrices (out of $d(d^2 - 1)$) such
that the difference between any two of them has 
nonzero determinant.  One can see that this property
guarantees (again with $\mu$ and $\nu$ nonzero) that
\begin{equation}  \label{sumto1}
\sum_S \delta_{S\mu, \nu} = 1.
\end{equation}
(If $\det(S_1 - S_2) \ne 0$, it follows
that for any nonzero $\alpha$, $(S_1 - S_2)\alpha \ne 0$, so that $S_1\alpha \ne S_2\alpha$.
Thus the $d^2 - 1$ distinct $S$ matrices must send $\alpha$
to the $d^2 - 1$ nonzero points.)
We can summarize Eqs.~(\ref{sumtod}) and (\ref{sumto1})
via our symbol ${\mathcal Z}$, which equals $d$ when
we are using the whole symplectic group and equals $1$
when we are using a minimal reconstructing set:
\begin{equation} \label{sumtoF}
\sum_S \delta_{S\mu, \nu} = {\mathcal Z}, \hspace{4mm} \mu,\nu \ne 0.
\end{equation}
\begin{widetext}
We now apply this fact to Eq.~(\ref{Ssum}).
If $\alpha$ equals $\alpha'$, the sum over $S$ is zero unless $\beta$ equals $\beta'$, in which case the sum is ${\mathcal Z}(d^2 - 1)$.
If $\alpha$ is not equal to $\alpha'$, the sum over $S$ is zero unless $\beta$ is not equal to $\beta'$, in which case the sum is ${\mathcal Z}$, in accordance with Eq.~(\ref{sumtoF}).  That is, we have
\begin{equation}
\begin{split} 
\hbox{LHS} &=
\frac{1}{{\mathcal Z}d^2}\left[ \delta_{\alpha\alpha'} \delta_{\beta\beta'} {\mathcal Z}(d^2 - 1) + (1 - \delta_{\alpha\alpha'}) (1 - \delta_{\beta\beta'}) {\mathcal Z} \right] \\
&= \delta_{\beta\beta'} \delta_{\alpha \alpha'} 
+\frac{1}{d^2} ( 1 - \delta_{\beta\beta'} - \delta_{\alpha\alpha'}),
\end{split} 
\end{equation}
which is what we wanted to prove.

%\begin{widetext}
\section{Distributing the $\Delta$'s}\label{distribute-Delta}
Here we prove a relationship that we call \textit{distributing the $\Delta$'s}. We will treat the case where a preparation, channel, and measurement are performed. We wish to 
show that
\begin{equation}
%\begin{split}
     \Delta \left[ \sum_{\beta,\alpha} R^{B'}(E|\beta) R^S_{\mathcal{E}}(\beta|\alpha) R^B(\alpha|{w}) \right] 
      =  \sum_{\beta,\alpha} \Delta R^{B'}(E|\beta) \Delta R^S_{\mathcal{E}}(\beta|\alpha) \Delta R^B(\alpha|{w}).
%\end{split}
\end{equation}
We start by rewriting the left-hand side using two applications of our rule for $\Delta$. This gives
%\begin{widetext}
\begin{equation}
    \begin{split}
    & \Delta \left[ \sum_{\beta,\alpha} R^{B'}(E|\beta) R^S_{\mathcal{E}}(\beta|\alpha) R^B(\alpha|{w}) \right] \\
    & \quad = \sum_{\beta,\alpha} R^{B'}(E|\beta) R^S_{\mathcal{E}}(\beta|\alpha) R^B(\alpha|{w}) - \frac{1}{d} \hbox{tr}(E)\\
    & \quad = \sum_{\beta,\alpha} \left( \Delta R^{B'}(E|\beta) + \frac{1}{d}\hbox{tr}(E)\right) \left( \Delta R^S_{\mathcal{E}}(\beta|\alpha) + \frac{1}{d^2}\right) \left( \Delta R^B(\alpha|{w}) + \frac{1}{d^2}\right) - \frac{1}{d} \hbox{tr}(E) .
\end{split}
\end{equation}
\end{widetext}
We multiply out the parenthetical factors and note that every term that contains exactly one or two factors of the form $\Delta R$ will be found to be zero once the summations over phase space are implemented. 
This is due to the following equations that hold true because the normalization of the $R$'s will always be equal to that of the $Q$'s:
% Eqs.~\eqref{DeltaDelta},~\eqref{recoverQ},~\eqref{Deltameas},
% This is due to the following two sets of equations: 
% \begin{eqnarray}
%     && \Delta Q_{w} (\gamma) = \sum_B \Delta R^B(\gamma | {w}),\\
%     && \Delta Q_{\mathcal{E}}(\alpha|\gamma) = \frac{1}{\mathcal{Z}}\sum_S \Delta R^S_\mathcal{E}(\alpha|\gamma),\\
%     && \Delta Q(E|\alpha)= \sum_{B'} \Delta R^{B'}(E|\alpha),
% \end{eqnarray}
\begin{equation}
    \begin{split}
    & \sum_\alpha \Delta R^B(\alpha|{w}) = 0,\\
    & \sum_\alpha \Delta R^S_\mathcal{E}(\beta|\alpha) = \sum_\beta \Delta R^S_\mathcal{E}(\beta | \alpha) = 0,\\
    & \sum_\beta \Delta R^{B'}(E|\beta) = 0.
\end{split}
\end{equation}
Any term with fewer than the maximal possible number of $\Delta R$'s will always have a factor of $\Delta R$ that is dependent on an unpaired phase space point having no match in the same term. 
For example, $\sum_{\beta,\alpha} \frac{1}{d^2} \Delta R^{B'}(E|\beta) \Delta R^S_\mathcal{E}(\beta|\alpha)$ has an unpaired $\alpha$ so the sum over $\alpha$ yields zero.
The two constant terms with values $\pm \frac{1}{d}\hbox{tr}(E)$ cancel and the relationship we sought follows. Similar reasoning leads to the analogous equations when there is no channel or multiple ones.

\section{Interpreting $R^B$ in terms of the
single-particle phase space} \label{RBRB1}
Our aim here is to interpret
$R^B(\alpha_1 |{w}_1)$, in which
$B$ is a striation in the large phase space
${\mathbb F}_d^2$, in terms of the smaller
phase space ${\mathbb F}_r^2$.

In Section \ref{primepower}, we have shown
that the $\Delta R^B(\alpha_1 | {w}_1)$'s are  
related to $\Delta Q(\alpha_1|{w}_1)$ by an analog of our standard 
formula (Eq.~(\ref{toomanyBs})).  We now show 
that for certain striations $B$, $\Delta R^B(\alpha_1 | {w}_1)$
is equal to $\Delta R^{B_1}(\alpha_1 | {w}_1)$
for a corresponding striation $B_1$ of the smaller 
phase space, and that for all other striations, 
$\Delta R^B(\alpha_1 | {w}_1)$ is zero.  

From the definitions of Section \ref{primepower}, we can write
\begin{equation}
R^B(\alpha_1 | w_1) = \frac{1}{d}\sum_{\alpha_2, \ldots, \alpha_n} \sum_{\beta \in \ell} Q(\alpha + \beta | w),
\end{equation}
where $\ell$ is the ray of the striation $B$, that is, the line in $B$ that passes through
the origin.
Doing the sum over $\alpha_2, \ldots, \alpha_n$, we get
\begin{equation}  \label{RB1}
R^B(\alpha_1 | w_1) = \frac{1}{d} \sum_{\beta \in \ell} Q(\alpha_1 + \beta_1 | w_1),
\end{equation} 
where $\beta_1$ is the ordered pair $(\beta_{q1}, \beta_{p1})$, the two components being
the coefficients of $e_1$ and $\tilde{e}_1$, respectively, in the expansions of $\beta_q$ 
and $\beta_p$.  

Let $m \in \mathbb{F}_d$ be the slope of the striation $B$
in the large phase space---assuming for now
that the slope is not infinite---and consider the equation of the ray $\ell$:
\begin{equation}  \label{longline}
\beta_p = m \beta_q.
\end{equation}
Taking the first-particle part of each side, we get
\begin{equation}
\beta_{p1}=\hbox{Tr}(e_1 \beta_p)
= \hbox{Tr}(e_1 m \beta_q)  .
\end{equation}
Now suppose $m$
is of the form $m = m_1 e_1^{-1}\tilde{e}_1$,
where $m_1 \in {\mathbb F}_r$.  Then
$\hbox{Tr}(e_1 m \beta_q)$ is equal to $m_1 \beta_{q1}$,
and we can write
\begin{equation}
\beta_{p1} = m_1 \beta_{q1} .
\end{equation}
This equation tells us that $\beta_1$ lies on the ray
in ${\mathbb F}_r^2$ with slope $m_1$.  Let us call this ray $\ell_1$ and the corresponding striation $B_1$.  Thus if $m$ has the form $m = m_1 e_1^{-1}\tilde{e}_1$, then as
$\beta$ ranges over the ray $\ell$ in ${\mathbb F}_d^2$, $\beta_1$ ranges over the ray $\ell_1$ in ${\mathbb F}_r^2$.  Moreover, it hits each point on this
ray exactly $r^{n-1}$ times, since the other coefficients in the expansion of $\beta_q$ take $r^{n-1}$ values for each value
of $\beta_{q1}$.     

Applying this result to Eq.~(\ref{RB1}), we have
\begin{equation}
\begin{split}
R^B(\alpha_1 | w_1) &= \frac{r^{n-1}}{d} \sum_{\beta_1\in\ell_1} Q(\alpha_1 + \beta_1 | w_1) \\
&= \frac{1}{r} \sum_{\beta_1\in\ell_1} Q(\alpha_1 + \beta_1 | w_1).  
\end{split}
\end{equation}
And this is exactly what we mean by $R^{B_1}(\alpha_1 | w_1)$ for the density matrix $w_1$
which is the partial trace of $w$.  

If the slope of $B$ is infinite, that is, if the ray $\ell$ is the solution to the equation
\begin{equation}
\beta_q = 0,
\end{equation}
then taking the first-particle part gives us
\begin{equation}
\beta_{q1} = 0.
\end{equation}
So again, the equation defines a ray in the smaller phase space, in this case the ray
with infinite slope, and we can continue the argument as in the preceding 
paragraph.  

The only other possibility is that 
$m = \sum_j m_j e_1^{-1}\tilde{e}_j$, with at least
one of the $m_j$'s, with $j\ne 1$, being nonzero.  
Let us suppose this is the case, and let $k\ne 1$
be such that $m_k$ is nonzero.  
Starting again with the equation (\ref{longline})
for the ray $\ell$ of slope $m$ in the large phase space, 
multiplying by $e_1$ and taking the trace, we 
get
\begin{equation}
\beta_{p1} = m_1 \beta_{q1} + m_k \beta_{qk} 
+ \sum_{j \notin \{1,k\}} m_j \beta_{qj} .
\end{equation}
We claim that as $\beta$ ranges over $\ell$, the point
$(\alpha_{q1},\alpha_{p1})$ in the small phase
space takes all the values in ${\mathbb F}_r^2$
and takes each such value the same number of times.
Suppose we first hold fixed every $\beta_{qj}$ with
$j\notin \{1,k\}$ and let just $\beta_{q1}$ and
$\beta_{qk}$ vary. Because $m_k$ is nonzero, we can
reach each value of $\beta_{p1}$ for each value of
$\beta_{q1}$: the $r^2$ values of $(\beta_{q1},\beta_{qk})$
map one-to-one to the $r^2$ values of 
$(\beta_{q1},\beta_{p1})$.  Moreover, this holds for every choice
of the $\beta_{qj}$'s with $j \notin \{1,k\}$.  So
as $\beta$ ranges over the whole ray, the ordered pair
$(\beta_{q1},\beta_{p1})$ hits every point in the 
small phase space exactly $r^{n-2}$ times.  Thus, for
a striation with a slope of this form (that is,
with some $m_j \ne 0$ with $j\ne 1$), the ray in 
the striation does not correspond to a ray in 
the small phase space.  Rather, it covers
the entire small phase space uniformly.  It follows that
the right-hand side of Eq.~(\ref{RB1}) does not depend on 
$\alpha_1$ at all; that is,
$R^B(\alpha_1 | {w}_1)$ is the 
uniform distribution.  So 
$\Delta R^B(\alpha_1 | {w}_1)$ is zero. 

Thus, in the right-hand side
of Eq.~(\ref{toomanyBs}), we can get a nonzero contribution
only for $r+1$ of the striations $B$, namely, those 
having slopes of
the form $m = m_1 e_1^{-1}\tilde{e}_1$ with $m_1 \in {\mathbb F}_r$ together with the striation having infinite slope.  
For these cases, we can replace $\Delta R^{B}(\alpha_1 | w_1)$ with $\Delta R^{B_1}(\alpha_1 | w_1)$.

\section{Constraints on the $R$ distributions}

Every possible quantum state can be made out of pure
states by convex combination.  And every possible
quantum operation can be made out of unitary channels,
together with the operations of appending an auxiliary system
and taking the partial trace.  In this
sense, if we give rules that define the possible
pure states and the possible unitary transformations,
then as long as we know how to do the 
required auxiliary 
manipulations, we have,
in effect, specified the complete set of quantum 
states and quantum operations.  Our aim in this
Appendix is to give the rules that define
pure states, unitary channels, and pure-state measurement outcomes when they
are expressed in terms of our $R$ distributions.
Since we know how to do a partial trace only when the dimension of the remaining system is
an odd prime power,
it is only for these cases that we claim to
have a full specification of quantum states, operations, 
and measurements.  Note that though we do not yet have a general way
of appending an auxiliary system (because we do not yet have a general
way of handing composite systems), when $r$ is an odd prime power we can use the partial trace to recognize
a product of pure states with dimensions $r$ and $r^2$, and this is a sufficient starting 
point for our present purposes.

To express the rules for pure states and unitary transformations, it is very helpful to
define the three-point structure function
\begin{equation}
\Gamma_{\alpha\beta\gamma} = \frac{1}{d}\hbox{tr}(A_\alpha
A_\beta A_\gamma),
\end{equation}
where the $A$'s are our phase-point operators. 
For the phase-point operators we have used in this
paper (from Refs.~\cite{Wootters,Klimov,Vourdas4}),
$\Gamma$ has the following simple form:
\begin{equation}
\Gamma_{\alpha\beta\gamma} = \frac{1}{d}\,
\omega^{-2\,\hbox{\scriptsize Tr}_{\mathbb{F}_d/\mathbb{F}_p}(\langle\alpha,\beta\rangle + \langle\beta,\gamma\rangle
+\langle\gamma,\alpha\rangle)},
\end{equation}
where $\langle\alpha,\beta\rangle$ is again the symplectic
product
\begin{equation}
\langle\alpha,\beta\rangle = \alpha_p\beta_q - \alpha_q\beta_p.
\end{equation}

In terms of $\Gamma$, we can recognize a 
pure state as a properly normalized Wigner
function that satisfies \cite{Wootters}
\begin{equation}  \label{purestatecondition}
Q(\alpha|{w}) = \sum_{\beta,\gamma}\Gamma_{\alpha\beta\gamma}
Q(\beta|{w})Q(\gamma|{w}).
\end{equation}
This condition comes from a familiar fact about
a pure-state density matrix ${w}$: ${w}^2 = {w}$.

We can also express in terms of $\Gamma$ the condition
for a unitary channel ${U}$.  The transition quasiprobabilities
$Q_{U}(\beta | \alpha)$, in addition to
satisfying the normalization condition 
$\sum_\beta Q_{U}(\beta | \alpha)=1$, must also 
preserve $\Gamma$ in the sense that
\begin{equation}  \label{unitarycondition}
\sum_{\alpha,\beta,\gamma}Q_{U}(\alpha'|\alpha)
Q_{U}(\beta'|\beta)Q_{U}(\gamma'|\gamma)
\Gamma_{\alpha\beta\gamma} = \Gamma_{\alpha'\beta'\gamma'}.
\end{equation}
Though the proof in Ref.~\cite{Wootters} that this condition is equivalent to
unitarity was meant to apply only to the Wigner function
defined in that paper, one finds that the proof applies
equally well to the closely related definition proposed
in Refs.~\cite{Klimov} and \cite{Vourdas4}, which we
have used when $d$ is a power of an odd prime.

Eqs.~(\ref{purestatecondition}) and (\ref{unitarycondition}) give us the conditions
for a pure state and a unitary channel, respectively,
in terms of the $Q$ distributions.  To convert these
equations into conditions on the $R$ distributions,
we use the relations between $Q$ and $R$ obtained
from Eqs.~(\ref{DeltaDelta}) and (\ref{recoverQagain}).
\begin{equation}  \label{RBtoQagain}
Q(\alpha|{w}) = \sum_B R^B(\alpha|{w}) - \frac{1}{d}
\end{equation}
and
\begin{equation}  \label{RStoQagain}
Q(\beta|\alpha) = \frac{1}{\mathcal Z}\sum_S R^S(\beta|\alpha) 
- \frac{d^2 - 2}{d^2}.
\end{equation}
So we can say that the desired constraints on
the $R$ distributions (for pure states and unitary
channels) are
Eqs.~(\ref{purestatecondition}) and (\ref{unitarycondition}) with each $Q$ distribution
replaced by the appropriate expression in Eq.~(\ref{RBtoQagain})
or (\ref{RStoQagain}).

As we note in the main text, these equations are not
particularly simple.  (Upon multiplying the factors out
and collecting similar terms, one does not achieve
any greater simplicity.)  But it is conceivable that
one can derive these equations from simpler principles
more aptly framed for our epistemically restricted
classical distributions.  

Finally, we need to address the functions 
$R^B(E|\alpha)$.  Again we start from the Wigner-function
formulation.  If the measurement outcome's POVM element
$E$ is of the form 
\begin{equation} \label{purePOVMelement}
E = |\psi\rangle\langle\psi|
\end{equation}
for
some state vector $|\psi\rangle$, the quasiprobability
function $Q(E|\alpha)$ must satisfy an equation analogous to Eq.~(\ref{purestatecondition}):
\begin{equation}  \label{puremeascondition}
Q(E|\alpha) = \frac{1}{d} \sum_{\beta,\gamma}\Gamma_{\alpha\beta\gamma}
Q(E|\beta)Q(E|\gamma).
\end{equation}
We convert this equation to a condition on the $R$'s 
via the analog of Eq.~(\ref{RBtoQagain}):
\begin{equation} \label{lastone}
Q(E|\alpha) = \sum_B R^B(E|\alpha) - 1.
%\frac{1}{d(d+1)}\sum_{B,\beta} R^B(E|\beta).
\end{equation}
  Suppose, then, that we have
a number of sets $\{R^B(E_j|\alpha)\}$ of classical distributions, each of which satisfies
Eqs.~(\ref{puremeascondition}) and (\ref{lastone}) (and thus represents
a POVM element of the form (\ref{purePOVMelement})).  Then any set of $R$'s
whose elements are of the form
\begin{equation} \label{generalPOVMelement}
R^B(E|\alpha) = \sum_j c_j R^B(E_j|\alpha),
\end{equation}
where each $c_j$ is a positive real number 
and $\sum_j c_j \le 1$, represents a legitimate
outcome of a measurement, typically not associated
with a pure state.  Moreover, any legitimate
outcome of a measurement can be obtained via
Eq.~(\ref{generalPOVMelement}), just as any mixed state can be
obtained as a convex combination of pure states.
(The difference here is that the $c_j$'s can
sum to a value {\em less} than one, since a POVM
element can have any normalization less than one.)
So Eqs.~(\ref{puremeascondition})--(\ref{generalPOVMelement}) 
together tell us what sets of $R$ functions are allowed in the
specification of a measurement outcome.

\end{document}